\RequirePackage{fix-cm}
\documentclass{article}
\usepackage{graphicx}
\usepackage{pgf,xcolor,tikz}
\usepackage{xfrac}
\usepackage{amsmath}
\usepackage{ulem}
\usepackage{textcomp}

\usepackage{amsfonts}
\usepackage{amssymb}
\usepackage{graphicx}
\usepackage{hyperref}
\usepackage{subfigure}
\begin{document}

\title{Dissipation in 2D degenerate gases with non-vanishing rest mass 
}


\author{A. R. M\'endez         \and
        A. L. Garc\'ia-Perciante \and  G. Chac\'on-Acosta 
}




\maketitle

\begin{abstract}\label{abs}
The complete set of transport coefficients for two dimensional relativistic degenerate gases is derived within a relaxation approximation in kinetic theory, by considering both the particle and energy frames. A thorough comparison between Marle and Anderson-Witting's models is carried out, pointing out the drawbacks of the former when compared both to the latter and to the full Boltzmann equation results in the non-degenerate limit. Such task is accomplished by solving the relativistic Uehling-Uhlenbeck equation, in both the particle and energy frames, in order to establish the constitutive equations for the heat flux and the Navier tensor together with analytical expressions for the transport coefficients in such representations. In particular, the temperature dependence of the thermal conductivity (associated with a generalized thermal force) and the bulk and shear viscosities are analyzed and compared within both models and with the non-degenerate, non-relativistic and ultra-relativistic limits.   
\end{abstract}

\section{Introduction}

Kinetic theory constitutes a powerful tool that allows for the establishment of transport equations for dilute systems. Moreover, its generalization to high temperature gases provides a consistent framework in which one can translate the relativistic corrections in the individual particles' dynamics and their interactions to measurable modifications to the transport properties of the fluid as a whole. Even though the foundations of relativistic kinetic theory have been established since the sixties (see for example \cite{Israel63}), the field is still very active and currently attracting further attention due to various applications in relativistic hydrodynamics, ranging from cosmology and astrophysics to experimental scenarios such as relativistic heavy-ion colliders; as examples of the variety of applications of relativistic gases, we cite Refs. \cite{Kelly73,apAC1,apAC2,apAC3,apHI1}. 

More recently, the interest in this context has also been focused on studying electron transport in two-dimensional materials such as graphene, which can be modeled as a two-dimensional relativistic Fermi gas \cite{apG1,apG2}. It is well known that in the tight-binding model for electrons moving on the surface, the chemical potential can be taken as zero at the neutral point; however, the average of this property can fluctuate due to the material's impurities. These variations are found mostly in the range from 5 to 56 meV depending on the experimental conditions \cite{Lucas,Svintsov}. Furthermore, it has been suggested that these small impurities can be modeled as local modifications in the viscosity within the hydrodynamic case \cite{Oettinger}. These facts also point to relativistic two-dimensional kinetic theory as an essential tool for the study and development of this new and promising material.

Mainly driven by the last application mentioned above, attention has been drawn to the determination of transport coefficients for relativistic quantum gases in two spatial dimensions. Furthermore, it has been shown in recent works that the Chapman-Enskog method yields good results for such
systems \cite{succi,Gabbana}, even though it has been argued that it can lead to unstable and non causal results in the first order (Navier-Stokes) regime (see for example Ref. \cite{HL}). Such method proposes, to first order in Knudsen's parameter, linear relations between dissipative fluxes and thermodynamic forces. The corresponding coupling parameters, namely the transport coefficients, can be approximated by simplifying the kinetic equations through the introduction of a relaxation model for the evolution of the distribution function.
The caveat of these approximate methods in the relativistic case is that, depending on the chosen space-time decomposition, which is in turn linked to the frame considered for the description of the thermodynamic properties of the system, one can consider two kinds of relaxation models. On the one hand, Marle's model is consistent with the so-called particle frame in which the time direction is aligned with the hydrodynamic velocity. On the other hand the Anderson-Witting approximation is the appropriate one if the energy frame is considered. In such description, the time direction is chosen such that no heat flux is present in the energy-momentum tensor. Although a debate on which of the approximations is more appropriate is beyond the scope of this work, we want to clearly state that the transport coefficients obtained with both models do not always lead to the same behavior, being Marle's model the one which seems to deviate the most from the expected results in the case of viscous dissipation. Indeed, the high temperature limit of the transport coefficients obtained in this case differ from those obtained from the Boltzmann equation, for specific cases by an inverse factor of the ratio between the rest mass and the temperature \cite{KremerLibro}. These flaws can be overcome by adapting the relaxation parameter included in the model to match the full Boltzmann equation's results as is proposed in Ref. \cite{MendezAIP2010} or considering the extended Marle model introduced by Takamoto et al. \cite{Takamoto}. It is worthwhile to mention that the difference in the transport coefficients may also be due to the expansion method carried out. Indeed, as can be verified in Ref. \cite{KremerLibro}, an implementation of a combined Chapman-Enskog and Maxwell iteration procedure can also lead to more accurate results. Establishing whether the inconsistency in Marle's results for the viscosity coefficients arises from the approximation method for the distribution function or from the collisional model itself is not an easy task which will be addressed in a separate work. In the present paper we consider both representations for the system in hand and discuss the results in a parallel fashion.

The transport coefficients in two-dimensional gases have been calculated for non-degenerate gases, both within the Anderson-Witting approximation \cite{succi} as well as using the complete Boltzmann equation for a hard disks model interaction \cite{Garcia-Mendez2019}. Moreover, the former has been successfully validated numerically in Refs. \cite{succi} and \cite{Gabbana}, while Ref. \cite{AM} presents a numerical analysis which exhibits consistency with the exact results reported in Ref. \cite{Garcia-Mendez2019}. The case of two-dimensional degenerate relativistic gases was first addressed in a study that validated the relativistic lattice Boltzmann's method, comparing it only with the theoretical results given by the Grad method \cite{Coelho}. 
Recently in Ref. \cite{nos19}, a relativistic fermion gas was studied in the particle frame using Marle's relaxation model. There, a constitutive equation for heat flux was established, driven by a generalized thermal force that includes both the temperature and chemical potential gradients. The present work shows the complete dissipation analysis as obtained by using the Chapman-Enskog method, for degenerate two-dimensional systems. The constitutive equations for the heat flux and the Navier tensor for non-zero rest mass bosons and fermions are estsblished, together with analytical expressions for the corresponding transport coefficients, namely the thermal conductivity, bulk, and shear viscosities. Massless particles should be treated separately. In such a case the dynamic pressure vanishes, there is no bulk viscosity and only the Anderson-Witting model can be used. This scenario will be addressed elsewhere.

The rest of this work is organized as follows. In Sect. 2 we state the Uehling-Uhlenbeck equation and briefly describe the Chapman-Enskog procedure by means of which the drift side of the kinetic equation is approximated to first order in the gradients. Section 3 introduces the particle frame and Marle's approximation for the collision term. The dissipative fluxes and corresponding transport coefficients are obtained. Similarly, in Sect. 4 the Anderson-Witting case is addressed, firstly describing the energy frame decomposition and then deriving the thermal conductivity and viscosities. The non-degenerate limit is analyzed in Sect. 5, where the Maxwell-Boltzmann case is also verified. Final remarks as well as comments on future work are included in Sect. 6. Appendices A-C contains the derivation of the local equilibrium (Euler) equations and the non-degenerate and non-relativistic limits, respectively.

\section{Basic setup of the problem}

\subsection{The Uehling-Uhlenbeck equation and conserved quantities}

As mentioned above, the system addressed in the present work is a dilute gas of degenerate, non-vanishing mass particles whose temperature can cover the complete range, even reaching sufficiently high values such that the relativistic parameter $\zeta=mc^{2}/kT$ may be negligible
(here $m$ is the particle's mass, $c$ the speed of light in vacuum,
$k$ Boltzmann's constant and $T$ the temperature). The evolution of the single particle distribution function $f(x^a,p^a)$ (that we write as $f$ in order to simplify the notation), in the absence of external fields, is thus given in general by the Uehling-Uhlenbeck namely \cite{KremerLibro,UU},
\begin{equation}
p^{a}\frac{\partial f}{\partial x^{a}}=J\left(ff'\right),\label{eq:Uhlebeck-1}
\end{equation}
with the collision kernel, $J\left(ff'\right)$, given by
\begin{equation}
J\left(ff'\right)=\int\left[\tilde{f}'f'\left(1-s\frac{h^{2}f}{g_{s}}\right) \left(1-s\frac{h^{2}\tilde{f}}{g_{s}}\right)\right.
\left. -\tilde{f}f\left(1-s\frac{ h^{2}f'}{g_{s}}\right)
\left(1-s\frac{h^{2}\tilde{f}'}{g_{s}}\right)\right] F\sigma d\Omega\frac{d^{2}\tilde{p}}{\tilde{p}_{0}},\label{2}
\end{equation}
where $s=1$ for fermions and $s=-1$ for bosons. In Eq. (\ref{eq:Uhlebeck-1})
the independent variables are $x^{a}$ and $p^{a}$, which stand for
the position and momentum 3-vectors, $x^{a}=\left(ct,\,x^{1},\,x^{2}\right)$
and $p^{a}=\left(p^{0},p^{1},p^{2}\right)$, with a flat space-time
metric $ds^{2}=c^{2}dt^{2}-dx^{2}-dy^{2}$. The right hand side includes
all variations in the occupation number of the phase space cells,
due to binary collisions, through the integral defined in Eq. (\ref{2}),
in which a prime denotes the corresponding distribution function evaluated
after a collision. $F$ stands for the invariant flux, $\sigma$ for
the 2D impact parameter and $\Omega$ is the solid angle. Also $h$
and $g_{s}$ are Planck's constant and the degeneracy factor respectively, and the abbreviations $\tilde{f}'\equiv f\left(x^a,\tilde{p}'^a\right)$, $\tilde{f}\equiv f\left(x^a,\tilde{p}^a\right)$ and $f'\equiv f\left(x^a,p'^a\right)$ were introduced.

The kinetic equation, due to the symmetry properties of the collision
kernel, is consistent with the conservation equations for the particle
flux and the energy momentum tensor:
\begin{equation}
\frac{\partial N^{a}}{\partial x^{a}}=0,\qquad\qquad\frac{\partial T^{ab}}{\partial x^{a}}=0,
\end{equation}
where these tensors are given by the following expressions
\begin{equation}
N^{a}=\int p^{a}f\left(\frac{d^{2}p}{p_{0}}c\right),\label{eq:f}
\end{equation}
\begin{equation}
T^{ab}=\int p^{a}p^{b}f\left(\frac{d^{2}p}{p_{0}}c\right)\textcolor{orange}{.}\label{f1}
\end{equation}
$N^{a}$ and $T^{ab}$ contain all fluxes, that is the ones corresponding
to the local equilibrium situation as well as the out of equilibrium,
dissipative ones. The explicit relation between these tensors and
such fluxes depends on the representation chosen and will be indicated
in Sections 3 and 4 for the particle and energy frames, respectively. 

\subsection{The Chapman-Enskog expansion}

Equation (\ref{eq:Uhlebeck-1}) is a non-linear integro-differential equation
which does not allow for an analytical solution. However, various
procedures have been proposed in order to obtain approximate solutions,
mostly close to equilibrium. In particular, the Chapman-Enskog method
has been shown to lead to accurate results in most regimes, and consists
in considering an expansion around the local equilibrium solution
namely \cite{Ch-E},
\begin{equation}
f=\sum_{i=0}^{\infty}f^{\left(i\right)}\textcolor{orange}{.}\label{eq:expanssion}
\end{equation}
Here $f^{\left(0\right)}$ corresponds to the equilibrium distribution
function which, for a relativistic degenerate gas, is given by
\begin{equation}
f^{(0)}=\frac{g_{s}}{h^{2}}\left(e^{\zeta\left(x-\alpha\right)}+s\right)^{-1}.\label{eq:f0Fermi-1-1}
\end{equation}
For convenience, we have introduced the following dimensionless parameters: $x=u^{a}p_{a}/mc^{2}$
and $\alpha=\mu_{e}/mc^{2}=\mu_{e}/kT\zeta$, with $\mu_{e}$ being the chemical potential, which takes different values for bosons and fermions. Also, the hydrodynamic velocity $u^{a}$ is given by
\begin{equation}
nu^{a}=\int p^{a}f^{\left(0\right)}\left(\frac{d^{2}p}{p_{0}}c\right),
\end{equation}
where $n$ is the particle density.

The subsequent corrections to $f^{\left(0\right)}$ are ordered according
to the so-called Knudsen parameter, which roughly indicates the relative
size of the microscopic scale (such as the particle's mean free path)
and the relevant macroscopic distance (characteristic scale of the
gradients). 
The Navier-Stokes regime, which corresponds to a first order in the
gradients theory, is recovered by truncating expansion (\ref{eq:expanssion})
to first order such that
\begin{equation}
f \approx f^{(0)}+f^{(1)},\label{eq:ce}
\end{equation}
where the first term on the right hand side contains all local equilibrium
information (Euler regime) while first order dissipation arises from
the correction $f^{(1)}$. Introducing such expansion, in Eq. (\ref{eq:Uhlebeck-1})
one is led to
\begin{equation}
J_{1}\left(ff'\right)=p^{a}\frac{\partial f^{\left(0\right)}}{\partial x^{a}}\label{3}
\end{equation}
where $J_{1}\left(ff'\right)$ denotes the collision term to first
order in Knudsen's parameter. 

The right hand side of Eq. (\ref{3}) can be readily expressed in
terms of the gradients of the state variables as follows
\begin{equation}
p^{a}\frac{\partial f^{\left(0\right)}}{\partial x^{a}}=\frac{g_{s}}{h^{2}}\frac{e^{\zeta\left(x-\alpha\right)}\zeta p^{a}}{\left(e^{\zeta\left(x-\alpha\right)}+s\right)^{2}} \left\{ \left(x-\alpha\right)\left[\frac{1}{T}\frac{\partial T}{\partial x^{a}}\right]+\alpha\left[\frac{1}{\mu_{e}}\frac{\partial\mu_{e}}{\partial x^{a}}\right]-\frac{p_{b}}{mc}\left[\frac{1}{c}\frac{\partial u^{b}}{\partial x^{a}}\right]\right\}, \label{eq:w}
\end{equation}
where the terms in square brackets correspond to the
gradients of the independent state variables chosen to describe the
system. 

Since linear irreversible thermodynamics requires dissipative fluxes,
arising from $f^{\left(1\right)}$ in this formalism, to be driven by the spatial gradients of the state variables, space and time need to be split using some decomposition. The standard decomposition in relativistic hydrodynamic is the so-called $d+1$ decomposition (for $d$ dimensions) in which one chooses some direction to indicate the time coordinate. Two representations are usually introduced in this sense: the particle frame in which the time direction is aligned with the hydrodynamic velocity, and the energy frame which is defined as the frame in which the heat flux is not present in the energy-momentum tensor. Independently of the frame chosen, the momentum vector can be decomposed in its time and space components in the following fashion:
\begin{equation}
p^{a}=p_{b}\left(\eta^{ab}-\frac{\text{e}^{a}\text{e}^{b}}{c^{2}}\right)+\left(\frac{\text{e}^{b}p_{b}}{c^{2}}\right)\text{e}^{a},\label{eq:4}
\end{equation}
where $\text{e}^{a}$ here denotes the chosen time-like vector for
the representation.

The first parenthesis on the right hand side of
Eq. (\ref{eq:4}) is often called spatial projector, and represents
the hyperplane orthogonal to the time direction at each point and
thus the spatial hypersurfaces of the corresponding space-time foliation.
However, as the separation of $\text{e}^{a}$ in the energy and particle
frames can be shown to be proportional to a first order flux, in both
cases the hydrodynamic velocity and its corresponding projector are
the relevant directions in (\ref{3}). Thus, since in local equilibrium
both the particle and energy frames coincide, one can consider the decomposition of relativistic momentum as follows,
\begin{equation}
p^{a}=p_{b}h^{ab}+mxu^{a},
\end{equation}
where $h^{ab}=\eta^{ab}-u^{a}u^{b}/c^{2}$ is the spatial projector with the hydrodynamic velocity as time-like vector.
Then, the right hand side of the Eq. (\ref{eq:w}), decomposes accordingly:
\begin{align}
p^{a}\frac{\partial f^{\left(0\right)}}{\partial x^{a}} & =\frac{g_{s}}{h^{2}}\frac{\zeta e^{\zeta\left(x-\alpha\right)}}{\left(e^{\zeta\left(x-\alpha\right)}+s\right)^{2}}\left\{ h^{ac}p_{c}\left(\left(x-\alpha\right)\frac{1}{T}\frac{\partial T}{\partial x^{a}}+\frac{\alpha}{\mu_{e}}\frac{\partial\mu_{e}}{\partial x^{a}}-\frac{p_{b}}{mc^{2}}\frac{\partial u^{b}}{\partial x^{a}}\right)\right.\nonumber \\
 & +\left.mx\left(\left(x-\alpha\right)\frac{u^{a}}{T}\frac{\partial T}{\partial x^{a}}+\frac{\alpha u^{a}}{\mu_{e}}\frac{\partial\mu_{e}}{\partial x^{a}}-p_{b}\frac{u^{a}}{mc^{2}}\frac{\partial u^{b}}{\partial x^{a}}\right)\right\}. \label{eq:5}
\end{align}

Notice that the second line in Eq. (\ref{eq:5}) includes the total proper time derivatives of the state variables. For consistency, and to guarantee the existence of the Chapman-Enskog solution, these terms need to be expressed in terms of the spatial gradients by introducing the previous order hydrodynamic equations. Such equations, which correspond to the local equilibrium regime, coincide in both frames and are given by
\begin{equation}
\left(\frac{u^{a}}{T}\frac{\partial T}{\partial x^{a}}\right)=\frac{1}{\zeta}\frac{\mathcal{I}_{2}^{s}I_{1}^{s}-\frac{1}{2}\mathcal{I}_{1}^{s}\left(3I_{2}^{s}-I_{0}^{s}\right)}{\mathcal{I}_{1}^{s}\mathcal{I}_{3}^{s}-\left(\mathcal{I}_{2}^{s}\right)^{2}}\frac{\partial u^{a}}{\partial x^{a}},\label{eq:6a}
\end{equation}
\begin{equation}
\left(\frac{\alpha u^{a}}{\mu_{e}}\frac{\partial\mu_{e}}{\partial x^{a}}\right)=-\frac{1}{\zeta\mathcal{I}_{1}^{s}}\left(I_{1}^{s}+\left(\mathcal{I}_{2}^{s}-\alpha\mathcal{I}_{1}^{s}\right)\frac{\mathcal{I}_{2}^{s}I_{1}^{s}-\frac{1}{2}\mathcal{I}_{1}^{s}\left(3I_{2}^{s}-I_{0}^{s}\right)}{\mathcal{I}_{1}^{s}\mathcal{I}_{3}^{s}-\left(\mathcal{I}_{2}^{s}\right)^{2}}\right)\frac{\partial u^{a}}{\partial x^{a}},\label{eq:7a}
\end{equation}
\begin{equation}
\left(\frac{u^{a}}{c^{2}}\frac{\partial u^{b}}{\partial x^{a}}\right)=h^{ab}\left[\left(1+\frac{2\alpha I_{1}^{s}}{I_{0}^{s}-3I_{2}^{s}}\right)\left(\frac{1}{T}\frac{\partial T}{\partial x^{a}}\right)-\frac{2\alpha I_{1}^{s}}{I_{0}^{s}-3I_{2}^{s}}\left(\frac{1}{\mu_{e}}\frac{\partial\mu_{e}}{\partial x^{a}}\right)\right],\label{eq:8a}
\end{equation}
where the integrals $I_{n}^{s}$ and $\mathcal{I}_{n}^{s}$ are defined
as follow:
\begin{equation}
I_{n}^{s}=\frac{1}{h^{2}}\int\frac{x^{n}}{e^{\zeta\left(x-\alpha\right)}+s}dx,\label{eq:s-1}
\end{equation}
\begin{equation}
\mathcal{I}_{n}^{s}=\frac{1}{h^{2}}\int\frac{e^{\zeta\left(x-\alpha\right)}x^{n}}{\left(e^{\zeta\left(x-\alpha\right)}+s\right)^{2}}dx.\label{s2}
\end{equation}

\noindent In Appendix A we have included a somewhat detailed derivation
of Eqs. (\ref{eq:6a})-(\ref{eq:8a}). Introducing Eqs. (\ref{eq:6a})-(\ref{eq:8a})
in Eq. (\ref{eq:5}) one is lead to the following expression
\begin{align}
p^{a}\frac{\partial f^{\left(0\right)}}{\partial x^{a}} & =\frac{g_{s}}{h^{2}}\frac{\zeta e^{\zeta\left(x-\alpha\right)}}{\left(e^{\zeta\left(x-\alpha\right)}+s\right)^{2}}\left\{ \frac{m}{\zeta}\left[x^{2}\left(\frac{\mathcal{I}_{2}^{s}I_{1}^{s}-\frac{1}{2}\mathcal{I}_{1}^{s}
\left(3I_{2}^{s}-I_{0}^{s}\right)}{\mathcal{I}_{1}^{s}\mathcal{I}_{3}^{s}-
\left(\mathcal{I}_{2}^{s}\right)^{2}}+\frac{\zeta}{2}\right)\right.\right.\nonumber \\
 & \left.-x\left(\frac{\mathcal{I}_{2}^{s}}{\mathcal{I}_{1}^{s}}\frac{\mathcal{I}_{2}^{s}I_{1}^{s}-
 \frac{1}{2}\mathcal{I}_{1}^{s}\left(3I_{2}^{s}-I_{0}^{s}\right)}{\mathcal{I}_{1}^{s}
 \mathcal{I}_{3}^{s}-\left(\mathcal{I}_{2}^{s}\right)^{2}}+\frac{I_{1}^{s}}{\mathcal{I}_{1}^{s}} \right)-\frac{\zeta}{2}\right]\left(\frac{\partial u^{a}}{\partial x^{a}}\right)\\ 
 & \left.-\frac{p^{c}p^{d}}{mc^{2}}\left(\mathring{\sigma}_{cd}\right)-
 \alpha\left(\frac{2I_{1}^{s}}{I_{0}^{s}-3I_{2}^{s}}x+1\right)
 \left(\frac{1}{T}\frac{\partial T}{\partial x^{a}}-\frac{1}{\mu_{e}}\frac{\partial\mu_{e}}{\partial x^{a}}\right)h_{b}^{a}p^{b}\right\}\label{pb}  
\end{align}
where $\mathring{\sigma}_{cd}$ is the traceless symmetric component
of $\left(\partial u^{b}/\partial x^{a}\right)$,
\begin{equation}
\mathring{\sigma}_{ab}=\frac{1}{2}h_{a}^{c}h_{b}^{d}
\left(\frac{\partial u_{c}}{\partial x^{d}}+\frac{\partial u_{d}}{\partial x^{c}}\right)-\frac{1}{2}h_{ab}\frac{\partial u^{c}}{\partial x^{c}},
\end{equation}
and we have used:
\begin{equation}
h^{ac}p_{c}p_{b}\frac{\partial u^{b}}{\partial x^{a}}=\left(p^{c}p^{d}\mathring{\sigma}_{cd}+\frac{1}{2}m^{2}c^{2}
\left(1-x^{2}\right)\frac{\partial u^{e}}{\partial x^{e}}\right).
\end{equation}
In Sections 3 and 4, the constitutive equations for the heat flux
and the viscous tensor are established by calculating such fluxes,
using the solutions for $f^{\left(1\right)}$ as obtained from Eq. (\ref{3}), considering Marle and Anderson-Witting's relaxation approximations
respectively.

\section{Transport coefficients in Marle's approximation}

\subsection{The particle frame}

The particle frame is characterized by the particle flux being a time-like
vector in the direction of the hydrodynamic velocity. In such a frame,
the particle flux and energy momentum tensor are given by \cite{Eckart}
\begin{equation}
N^{a}=nu^{a},\label{n-1}
\end{equation}
and 
\begin{equation}
T^{ab}=\frac{n\epsilon}{c^{2}}u^{a}u^{b}-\mathrm{p}h^{ab}+\Pi^{ab}+\frac{1}{c^{2}}q^{a}u^{b}+\frac{1}{c^{2}}u^{a}q^{b},\label{t-1}
\end{equation}
where the equilibrium quantities introduced can be identified as:
\[
n=\frac{1}{c^{2}}\int u_{a}p^{a}f^{\left(0\right)}\left(\frac{d^{2}p}{p_{0}}c\right)\qquad\text{particle number density},
\]
\[
n\epsilon=\frac{1}{c^2}\int\left(u_{a}p^{a}\right)^{2}f^{\left(0\right)}\left(\frac{d^{2}p}{p_{0}}c\right)\qquad\text{internal energy density},
\]
\[
\mathrm{p}=\frac{h_{ab}}{2}\int p^{a}p^{b}f^{\left(0\right)}\left(\frac{d^{2}p}{p_{0}}c\right)\qquad\text{hydrostatic pressure.}
\]
The previous quantities can be written in terms of the
integrals defined in Eq. (\ref{eq:s-1}) as follows
\[
n=2\pi m^{2}c^{2}g_{s}I_{1}^{s},
\]
\[
n\epsilon=2\pi m^{3}c^{4}g_{s}I_{2}^{s},
\]
\[
\mathrm{p}=\pi m^{3}c^{4}g_{s}\left(I_{2}^{s}-I_{0}^{s}\right).
\]
For the dissipative terms we obtain the following expressions:
\begin{equation}
q^{a}=2\pi mc^{2}u_{b}h_{c}^{a}\int p^{b}p^{c}f^{\left(1\right)}dx,\label{j}
\end{equation}
\begin{equation}
\Pi^{ab}=2\pi mc^{2}h_{c}^{a}h_{d}^{b}\int p^{c}p^{d}f^{\left(1\right)}dx.\label{ji}
\end{equation}
Notice that the change of variable
\[
\left(\frac{u_{a}p^{a}}{mc^{2}}\right)=x\qquad\left(\frac{d^{2}p}{p_{0}}c\right)=2\pi mc^{2}dx,
\]
has been introduced in order to simplify the notation. In the following
subsection, the first order correction to the local equilibrium distribution
function is established within Marle's relaxation approximation.

\subsection{Marle's approximation}

In order to simplify the solution to Eq. (\ref{3}), the simplest
approach is to substitute the integral collision kernel by a relaxation
term. Marle's proposal consists in considering \cite{KremerLibro,Marle}
\begin{equation}
p^{a}\frac{\partial f_{\text{M}}}{\partial x^{a}}=-\frac{m}{\tau_{{\rm M}}}\left(f_{\text{M}}-f^{\left(0\right)}\right),\label{marle}
\end{equation}
where $\tau_{\text{M}}$ is a relaxation parameter and the subscript
${\rm M}$ stands for Marle's solution. Introducing the Chapman-Enskog
expansion (Eq. (\ref{eq:ce})) one can readily obtain an expression
for $f_{\text{M}}^{\left(1\right)}$ which we write as $f_{\text{M}}^{\left(1\right)}=f_{{\rm \text{M}I}}^{\left(1\right)}+f_{{\rm \text{M}II}}^{\left(1\right)}$
where
\begin{align}
f_{{\rm \text{M}I}}^{\left(1\right)} & =\alpha\frac{\tau_{{\rm M}}}{m}\frac{g_{s}}{h^{2}}\frac{\zeta e^{\zeta\left(x-\alpha\right)}}{\left(e^{\zeta\left(x-\alpha\right)}+s\right)^{2}}\left(\frac{2I_{1}^{s}}{I_{0}^{s}-3I_{2}^{s}}x+1\right)\left(\frac{1}{T}\frac{\partial T}{\partial x^{a}}-\frac{1}{\mu_{e}}\frac{\partial\mu_{e}}{\partial x^{a}}\right)h_{b}^{a}p^{b},\label{fm1}
\end{align}
\begin{eqnarray}
f_{{\rm \text{M}II}}^{\left(1\right)} & = &\tau_{{\rm M}}\frac{g_{s}}{h^{2}}\frac{e^{\zeta\left(x-\alpha\right)}}{\left(e^{\zeta
\left(x-\alpha\right)}+s\right)^{2}}\left\{-\left[x^{2}\left(\frac{\mathcal{I}_{2}^{s}I_{1}^{s}-\frac{1}{2}\mathcal{I}_{1}^{s}
\left(3I_{2}^{s}-I_{0}^{s}\right)}{\mathcal{I}_{1}^{s}\mathcal{I}_{3}^{s}-\left(\mathcal{I}_{2}^{s}\right)^{2}}+\frac{\zeta}{2}\right)
\right.\right. \\ \label{w}
 &  &\left.\left.-\left(\frac{\mathcal{I}_{2}^{s}}{\mathcal{I}_{1}^{s}}\frac{\mathcal{I}_{2}^{s}I_{1}^{s}-\frac{1}{2}
 \mathcal{I}_{1}^{s}\left(3I_{2}^{s}-I_{0}^{s}\right)}{\mathcal{I}_{1}^{s}\mathcal{I}_{3}^{s}
 -\left(\mathcal{I}_{2}^{s}\right)^{2}}+\frac{I_{1}^{s}}{\mathcal{I}_{1}^{s}}\right)x-\frac{\zeta}{2}\right]
 \frac{\partial u^{a}}{\partial x^{a}}+\left(\frac{\zeta}{m^{2}c^{2}}p^{a}p^{b}\mathring{\sigma}_{ab}\right)\right\},\nonumber
\end{eqnarray}
correspond to the thermal and viscous parts respectively.

\subsection{Thermal dissipation}

To obtain a constitutive equation for the heat flux, we rewrite the thermodynamic force appearing in Eq. (\ref{fm1}), which is in terms of the temperature and chemical potential gradients, by the usual one which contains the temperature and pressure gradients as follows
\begin{equation}
\frac{\partial T}{\partial x^{d}}-\frac{T}{\mu_{e}}\frac{\partial\mu_{e}}{\partial x^{d}}=\frac{h_{e}}{\mu_{e}}\left(\frac{\partial T}{\partial x^{d}}-\frac{T}{nh_{e}}\frac{\partial\text{p}}{\partial x^{d}}\right),\label{eq:force}
\end{equation}
where $h_{e}=\epsilon+{\rm  p}/n$ is the enthalpy. The substitution
of Eqs. (\ref{fm1}) and (\ref{eq:force}) in Eq. (\ref{j}) leads
to
\begin{align}
q^{a}&=2\pi k\tau_{{\rm M}}\frac{g_{s}}{h^{2}}c^{2}\zeta^{2}\left(\frac{\partial T}{\partial x^{b}}-\frac{T}{nh_{e}}\frac{\partial\text{p}}{\partial x^{b}}\right)\left(\frac{3I_{2}^{s}-I_{0}^{s}}{2I_{1}^{s}}\right)h_{c}^{a}h_{d}^{b}\\
& \times \int p^{c}p^{d}\frac{e^{\zeta\left(x-\alpha\right)}}{\left(e^{\zeta\left(x-\alpha\right)}+s\right)^{2}}\left(\frac{2I_{1}^{s}}{I_{0}^{s}-3I_{2}^{s}}x^{2}+x\right)dx.\nonumber
\end{align}
This expression can be simplified by using the following relation
(see Appendix A in Ref. \cite{Garcia-Mendez2019})
\begin{equation}
h_{c}^{a}h_{d}^{b}\int A\left(x\right)p^{c}p^{d}dx=\frac{m^{2}c^{2}}{2}h^{ab}\int A\left(x\right)\left(1-x^{2}\right)dx,\label{d-1}
\end{equation}
by means of which one can obtain:
\[
q^{a}=\lambda h^{ab}\left(\frac{\partial T}{\partial x^{b}}-\frac{T}{nh_{e}}\frac{\partial\text{p}}{\partial x^{b}}\right),
\]
where the transport coefficient associated with heat dissipation is
given by
\begin{equation}
\lambda_{\text{M}}=\tau_{{\rm M}}\pi kg_{s}m^{2}c^{4}\zeta^{2}\left[\frac{\left(3I_{2}^{s}-I_{0}^{s}\right)}{2I_{1}^{s}}\left(\mathcal{I}_{1}^{s}-\mathcal{I}_{3}^{s}\right)+\left(\mathcal{I}_{4}^{s}-\mathcal{I}_{2}^{s}\right)\right].\label{eq:lambda-Marle}
\end{equation}
Expression (\ref{eq:lambda-Marle}) was already obtained in Ref. \cite{nos19}
and is here included for completeness. The reader can consult more
details in such reference. Figure \ref{fig:Marle} shows the dependence
of this transport coefficient with the relativistic parameter $\zeta$
for a boson ($s=-1$) and a fermion system ($s=1$), for different
values of $\alpha=\mu_e/kT\zeta$.

\subsection{Viscous dissipation}

In order to explore viscous dissipation, the component of the non-equilibrium
contribution to $f$ which depends on the derivatives of the hydrodynamic
velocity is considered, i. e. Eq. (\ref{w}). In particular, one can
write the corresponding constitutive equation as
\begin{equation}
\Pi_{cd}=\eta\mathring{\sigma}_{cd}+\mu h_{cd}\frac{\partial u^{a}}{\partial x^{a}},\label{x-1}
\end{equation}
where the first term corresponds to the shear component of stress,
being $\eta$ the shear viscosity, and the second term to the compression
stresses, where $\mu$ is the bulk viscosity coefficient.

Substitution of Eq. (\ref{w}) in Eq. (\ref{ji}) leads to
\begin{eqnarray}
\Pi^{ab} & = & 2\pi mc^{2}\tau_{{\rm M}}\frac{g_{s}}{h^{2}}\left\{ \frac{\zeta}{m^{2}c^{2}}\mathring{\sigma}_{ef}h_{c}^{a}h_{d}^{b}
\int p^{c}p^{d}p^{e}p^{f}\frac{e^{\zeta\left(x-\alpha\right)}}{\left(e^{\zeta\left(x-\alpha\right)}+s\right)^{2}}dx\right. \nonumber \\
 & &-\frac{\partial u^{e}}{\partial x^{e}}h_{c}^{a}h_{d}^{b}\int p^{c}p^{d}\frac{e^{\zeta\left(x-\alpha\right)}}{\left(e^{\zeta
 \left(x-\alpha\right)}+s\right)^{2}}\times\left[x^{2}\left(\frac{\mathcal{I}_{2}^{s}I_{1}^{s}-\frac{1}{2}\mathcal{I}_{1}^{s}
 \left(3I_{2}^{s}-I_{0}^{s}\right)}{\mathcal{I}_{1}^{s}\mathcal{I}_{3}^{s}-\left(\mathcal{I}_{2}^{s}\right)^{2}}
 +\frac{\zeta}{2}\right)\right.\nonumber \\ 
 & &\left.\left.-\left(\frac{\mathcal{I}_{2}^{s}}{\mathcal{I}_{1}^{s}}\frac{\mathcal{I}_{2}^{s}I_{1}^{s}-\frac{1}{2}
 \mathcal{I}_{1}^{s}\left(3I_{2}^{s}-I_{0}^{s}\right)}{\mathcal{I}_{1}^{s}\mathcal{I}_{3}^{s}-\left(\mathcal{I}_{2}^{s}
 \right)^{2}}+\frac{I_{1}^{s}}{\mathcal{I}_{1}^{s}}\right)x-\frac{\zeta}{2}\right]dx\right\}. \label{w-1}
\end{eqnarray}
The second integral can be readily simplified by using Eq. (\ref{d-1})
while for the first one, use can be made of
\[
h_{c}^{a}h_{d}^{b}\int A\left(x\right)p^{c}p^{d}p^{e}p^{f}dx=\frac{m^{4}c^{4}}{8}\left(h^{ab}h^{ef}+h^{ae}h^{bf}+h^{af}h^{be}\right)\int A\left(x\right)\left(1-x^{2}\right)^{2}dx,
\]
together with
\[
\mathring{\sigma}_{ef}\left(h^{cd}h^{ef}+h^{ce}h^{df}+h^{cf}h^{de}\right)=2\mathring{\sigma}^{cd},
\]
which follows from $\mathring{\sigma}_{ef}$ being traceless and orthogonal
to the hydrodynamic velocity. Introducing such expressions, Eq. (\ref{w-1})
can be written as Eq. (\ref{x-1}) where
\begin{equation}
\eta_{{\rm M}}=\tau_{{\rm M}}\frac{\pi}{2}g_{s}m^{3}c^{4}\zeta\left(\mathcal{I}_{4}^{s}-2\mathcal{I}_{2}^{s}+\mathcal{I}_{0}^{s}\right),\label{5-1}
\end{equation}
is the shear viscosity, and
\begin{eqnarray}\label{6-1}
\mu_{{\rm M}}&=&\tau_{{\rm M}}\frac{\pi}{2}g_{s}m^{3}c^{4}\zeta\Biggl\{ \left(\mathcal{I}_{4}^{s}-2\mathcal{I}_{2}^{s}+\mathcal{I}_{0}^{s}\right)
\Biggr. \\ 
&+ & \left. \frac{2}{\zeta\mathcal{I}_{1}^{s}}
\left[\frac{\mathcal{I}_{2}^{s}I_{1}^{s}
-\frac{1}{2}\mathcal{I}_{1}^{s}
\left(3I_{2}^{s}-I_{0}^{s}\right)}{\left(\mathcal{I}_{1}^{s}
\mathcal{I}_{3}^{s}-\left(\mathcal{I}_{2}^{s}\right)^{2}\right)}
\left(\mathcal{I}_{1}^{s}
\mathcal{I}_{4}^{s}-\mathcal{I}_{3}^{s}\mathcal{I}_{2}^{s}\right)-I_{1}^{s}
\left(\mathcal{I}_{3}^{s}-\mathcal{I}_{1}^{s}\right)\right]\right\}, \nonumber
\end{eqnarray}
the bulk viscosity. 

\begin{figure}
\begin{centering}
\includegraphics[width=6.1cm]{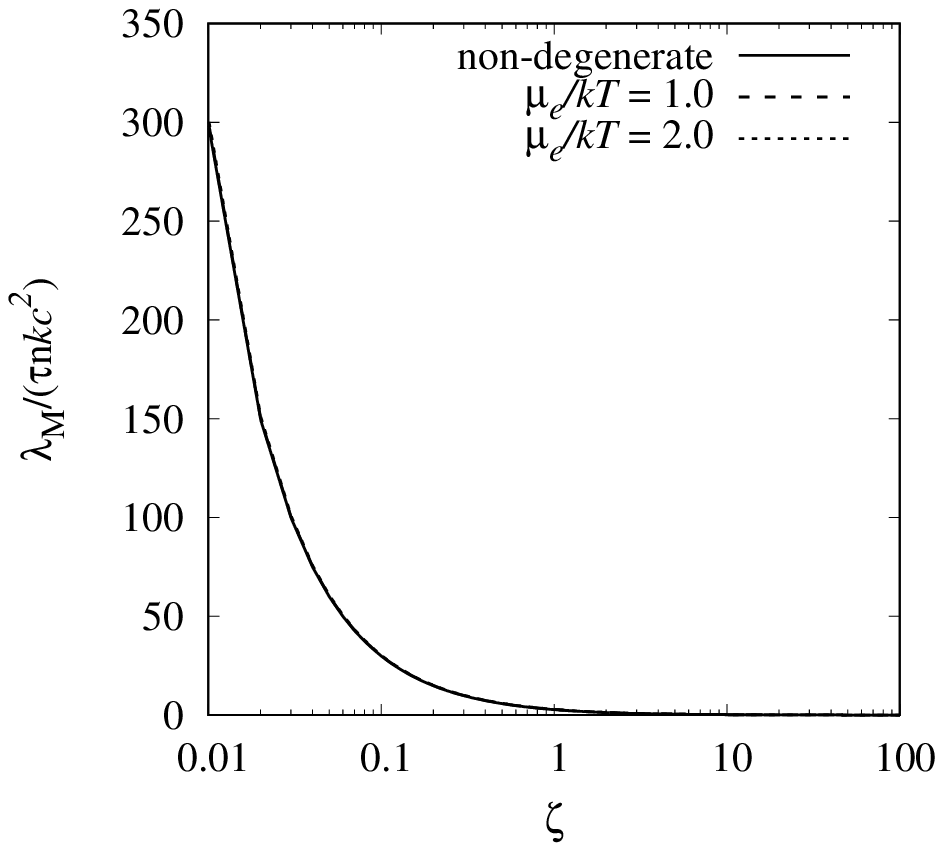}\includegraphics[width=6.1cm]{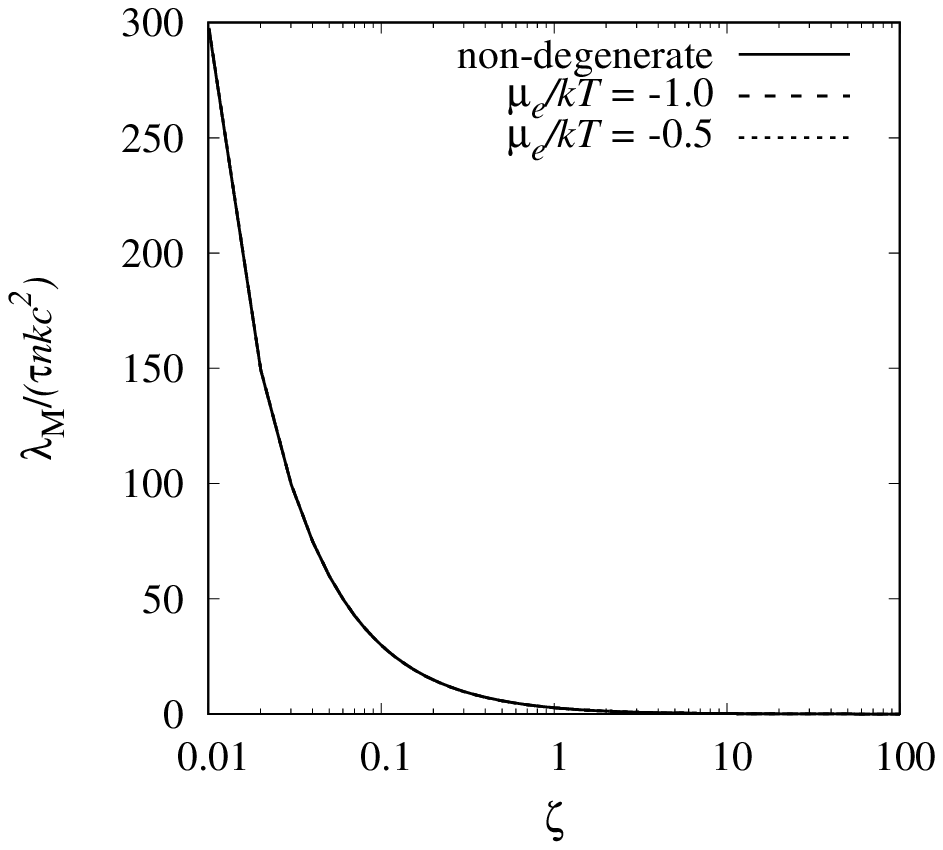}
\par\end{centering}
\begin{centering}
\includegraphics[width=6.1cm]{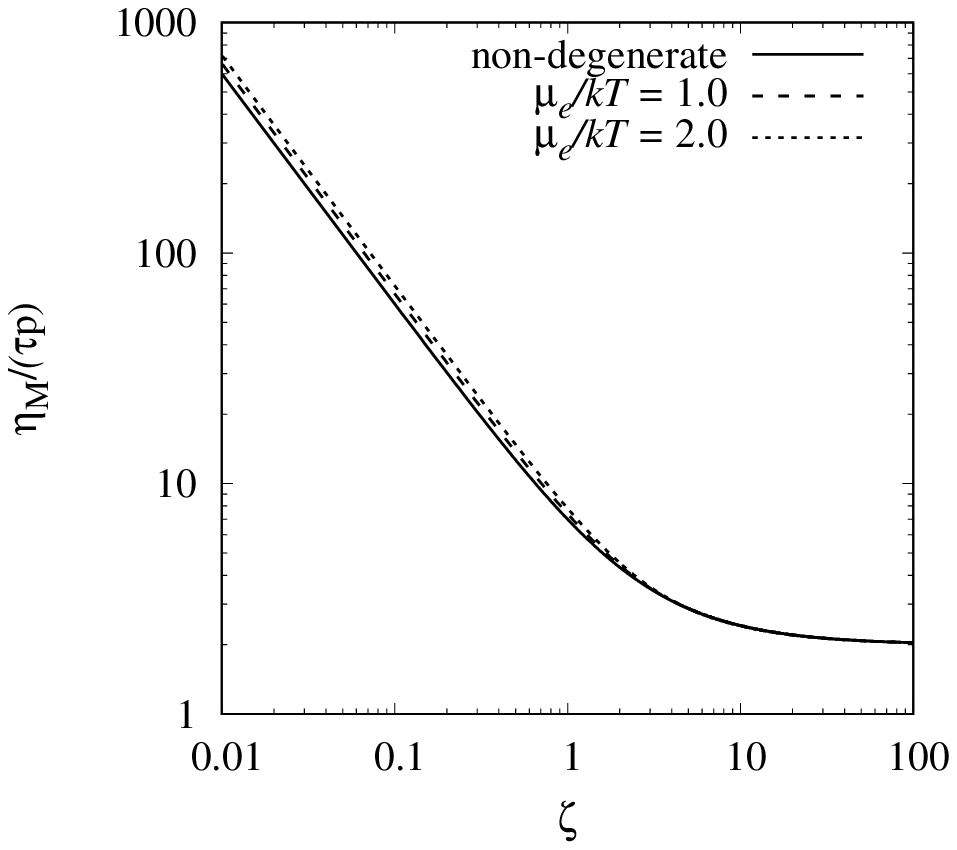}\includegraphics[width=6.1cm]{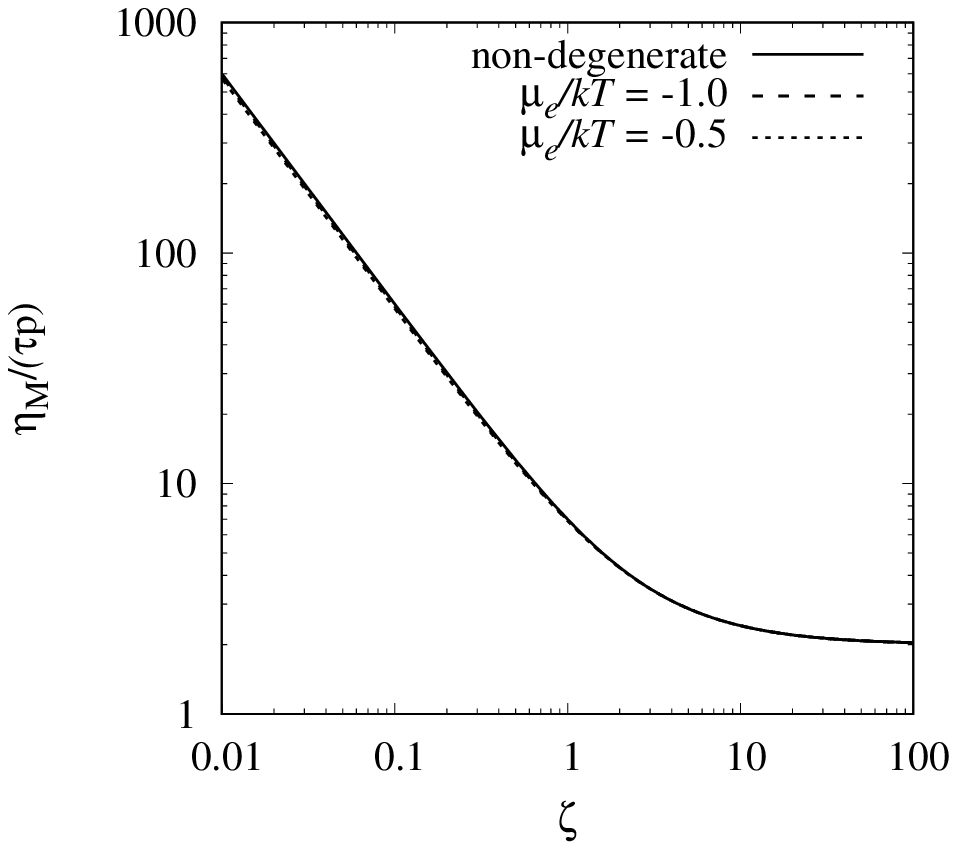}
\par\end{centering}
\begin{centering}
\includegraphics[width=6.1cm]{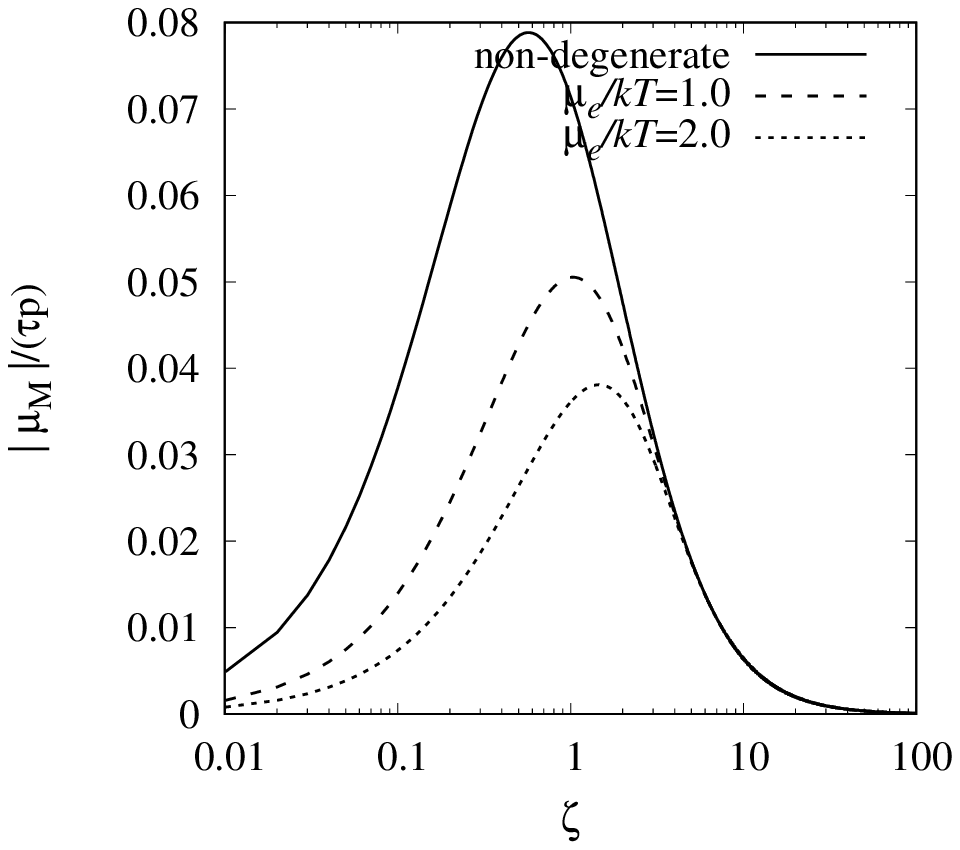}\includegraphics[width=6.1cm]{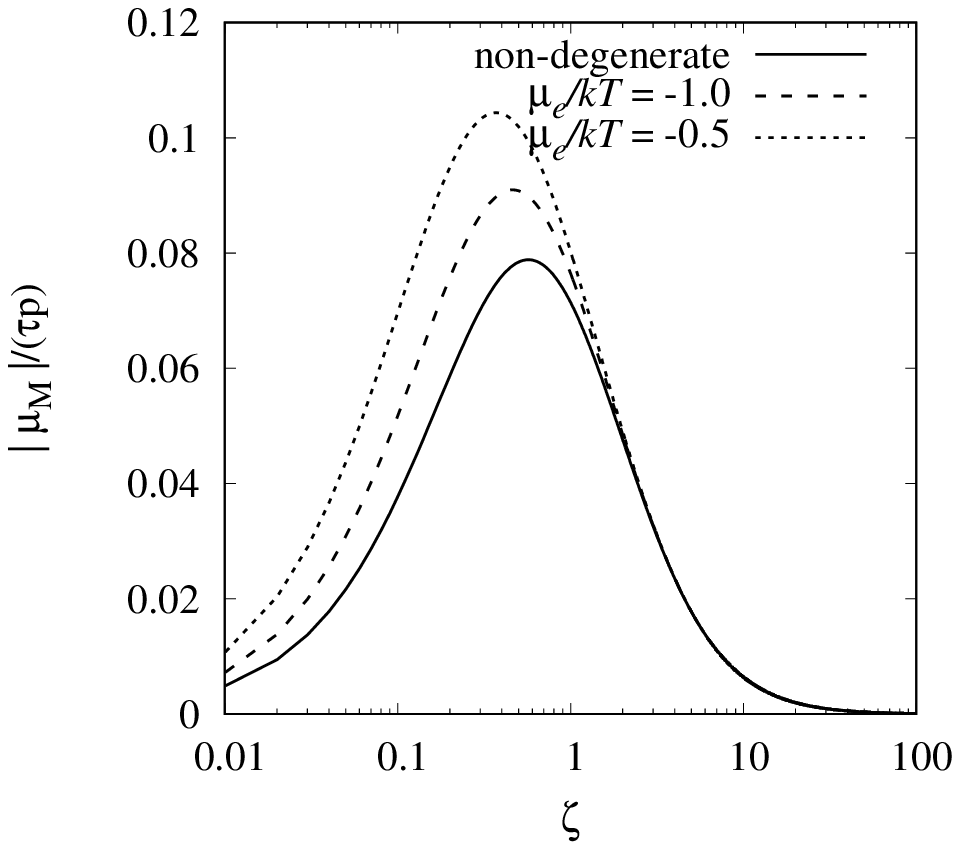}
\par\end{centering}
\caption{\label{fig:Marle}The transport coefficients obtained with Marle's
model: the thermal conductivity for fermions (top-left) and bosons
(top-right), the shear viscosity for fermions (center-left) and bosons
(center-right), and the bulk viscosity for fermions (bottom-left)
and bosons (bottom-right). 
}
\end{figure}

Figure \ref{fig:Marle} shows the dependence of
these coefficients with $\zeta$ for both types of particles considering different values of the chemical potential, which in the case of fermions can take any value since the occupation number in a given state can be only zero or one (at zero temperature), according to Pauli's exclusion principle. For bosons, when the chemical potential tends to the ground state, the occupation number diverges. Thus if $\mu_e $ is greater than the lowest energy state, it would have a negative occupation number. This is solved by imposing a negative chemical potential and setting its upper limit to zero.

In Fig. \ref{fig:Marle} both the thermal conductivity and shear viscosity for the Marle model exhibit a decreasing behaviour in $\zeta$. One can also note that both curves are practically overlapped by their non-degenerate counterpart, only showing a slight deviation for higher values of the chemical potential. On the other hand, the bulk viscosity presents a maximum value and vanishes in the extreme limits. This phenomenon, which indicates that the bulk viscosity is only relevant in intermediate regimes, is also exhibited in the non-degenerate case and in three dimensional scenarios.

At this point, we would like to draw attention to the fact that the bulk viscosity within this model yields negative values (Fig. 1 shows its absolute value). In Ref. \cite{Garcia-Mendez2019}, where the non-degenerate non-relativistic 2D case is studied using the complete Boltzmann equation, it is shown that the bulk viscosity coefficient is positive. Although a more detailed analysis of this point will be undertaken elsewhere, one can suspect that the negative sign in Marle's bulk viscosity could have some effect on the entropy production, i.e., the negative sign could allow, in some cases, the entropy production to cancel or even be negative, in contradiction with the second law of thermodynamics.

Also, notice that the shear viscosity coefficient, in this case, follows a different trend from the one reported in the literature \cite{succi,Gabbana,Garcia-Mendez2019}. This point is discussed in the final section of the present work, where the comparison with the Anderson-Witting model is addressed.

\section{Transport coefficients in Anderson-Witting's approximation}

\subsection{The energy frame}

Even though the particle frame may seem like the natural choice to
describe a relativistic fluid, which is also analogous in structure to non-relativistic hydrodynamics, there exists an alternate proposal
for the space-time decomposition due to Landau and Lifshitz, which
is usually referred to as the energy frame \cite{KremerLibro,L=000026L}.
In such a frame, the corresponding particle flux has an additional space-like component in a non-equilibrium situation, while the stress-energy tensor only features viscous dissipation and no heat flux. As a consequence, by choosing a frame in which 
\begin{equation}
T^{ab}=\frac{n\epsilon}{c^{2}}U^{a}U^{b}-\mathrm{p}h^{ab}+\Pi^{ab},\label{t-1-1}
\end{equation}
one requires a particle flow as
\begin{equation}
N^{a}=nU^{a}+\mathcal{J}^{a},\label{n-1-1}
\end{equation}
where $\mathcal{J}^{a}U_{a}=0$ and $nU^{a}+\mathcal{J}^{a}=nu^{a}$.
It can be shown that close to equilibrium, the dissipative current
$\mathcal{J}^{a}$ is proportional to the heat flux \cite{KremerLibro}, more precisely
\[
\mathcal{J}^{a}=-\frac{nq^{a}}{n\epsilon+p},
\]
such that in this representation, the heat flux is given by
\begin{equation}
q^{a}=-2\pi mc^{2}\left(\epsilon+\frac{p}{n}\right)h^{ab}\int p_{b}f^{\left(1\right)}dx,\label{eq:d}
\end{equation}
and the stress energy tensor retains its form, Eq. (\ref{ji}).
Notice that, even though the projectors involved in both expressions, Eqs. (\ref{ji}) and (\ref{eq:d}), should be orthogonal to $U^{a}$
in this representation, in order to keep the first order regime ($f^{\left(1\right)}$
is already first order in the gradients), the terms of the spatial
projector $\left(\eta^{ab}-\frac{1}{c^{2}}U^{a}U^{b}\right)$ are to be kept are simply as $\left(\eta^{ab}-\frac{1}{c^{2}}u^{a}u^{b}\right)=h^{ab}$.

\subsection{Anderson-Witting's approximation}

In the representation described above, the relaxation approximation
for the collision kernel is given by \cite{KremerLibro,AW1,AW2}
\begin{equation}
p^{a}\frac{\partial f_{\text{AW}}}{\partial x^{a}}=-\frac{mx}{\tau_{AW}}\left(f_{\text{AW}}-f^{\left(0\right)}\right)\label{aw}
\end{equation}
where as before, $\tau_{AW}$ is a relaxation parameter, in principle different from the one featured in Marle's case. The ${\rm AW}$ subscript denotes this difference indicating the Anderson-Witting frame.

Following the same steps as in the previous
section one is lead to the first order components of non-equilibrium
contribution to the distribution function. If we separate, as in Subsect. 3.2, the viscous and thermal contributions of $f_{{\rm \text{AW}}}^{\left(1\right)}$
in the form $f_{{\rm \text{AW}}}^{\left(1\right)}=f_{{\rm \text{AWI}}}^{\left(1\right)}+f_{{\rm \text{AWII}}}^{\left(1\right)}$,
and since the difference in the models is the extra $x$ factor
in the collision term (\ref{aw}), one obtains $f_{{\rm \text{AW}}i}^{\left(1\right)}=\left(\tau_{{\rm M}}/\tau_{{\rm AW}}\right)x^{-1}f_{{\rm M}i}^{\left(1\right)}$ for
$i=$I, II.
  Values for the relaxation parameters' ratio as well as models for the particular value for each one of them have been proposed such that the results obtained in the particle frame are in agreement with those obtained by the full relativistic Boltzmann equation and by the model in the energy frame. For a thorough discussion of the value of the relaxation parameter in both models, the reader can resort to Refs. \cite{MendezAIP2010,Takamoto}.

\subsection{Thermal dissipation}

Substitution of $f_{{\rm \text{AWI}}}^{\left(1\right)}$ in the particle
flux projection indicated in Eq. (\ref{eq:d}) yields
\[
q^{a}=-2\pi k\tau_{\text{AW}}\frac{g_{s}}{h^{2}}\frac{h_{e}^{2}}{m^{2}c^{2}}\zeta^{2}\left(\frac{\partial T}{\partial x^{d}}-\frac{T}{nh_{e}}\frac{\partial\text{p}}{\partial x^{d}}\right)h_{b}^{a}h_{c}^{d}\int p^{c}p^{b}\frac{e^{\zeta\left(x-\alpha\right)}}{\left(e^{\zeta\left(x-\alpha\right)}+s\right)^{2}}\left(\frac{2I_{1}^{s}}{I_{0}^{s}-3I_{2}^{s}}+\frac{1}{x}\right)dx,
\]
in which, the use of Eq. (\ref{d-1}) leads to
\[
q^{a}=-\pi\tau_{\text{AW}}g_{s}kh_{e}^{2}\zeta^{2}h^{ab}\left(\frac{\partial T}{\partial x^{b}}-\frac{T}{nh_{e}}\frac{\partial\text{p}}{\partial x^{b}}\right)\left(\frac{2I_{1}^{s}}{I_{0}^{s}-3I_{2}^{s}}\left(\mathcal{I}_{0}^{s}-\mathcal{I}_{2}^{2}\right)+\left(\mathcal{I}_{-1}^{s}-\mathcal{I}_{1}^{s}\right)\right).
\]
Thus, the coefficient related to thermal conduction in this approximation
is given by
\begin{equation}
\lambda_{\text{AW}}=\pi\tau_{\text{AW}}g_{s}km^{2}c^{4}\zeta^{2}\left(\frac{3I_{2}^{s}-I_{0}^{s}}{2I_{1}^{s}}\right)\left(\left(\mathcal{I}_{0}^{s}-\mathcal{I}_{2}^{s}\right)+\left(\frac{3I_{2}^{s}-I_{0}^{s}}{2I_{1}^{s}}\right)\left(\mathcal{I}_{1}^{s}-\mathcal{I}_{-1}^{s}\right)\right),\label{lambdaAW}
\end{equation}
which is shown in Fig. \ref{fig:AW-1}.

\subsection{Viscous dissipation}

\begin{figure}
\begin{centering}
\includegraphics[width=6.1cm]{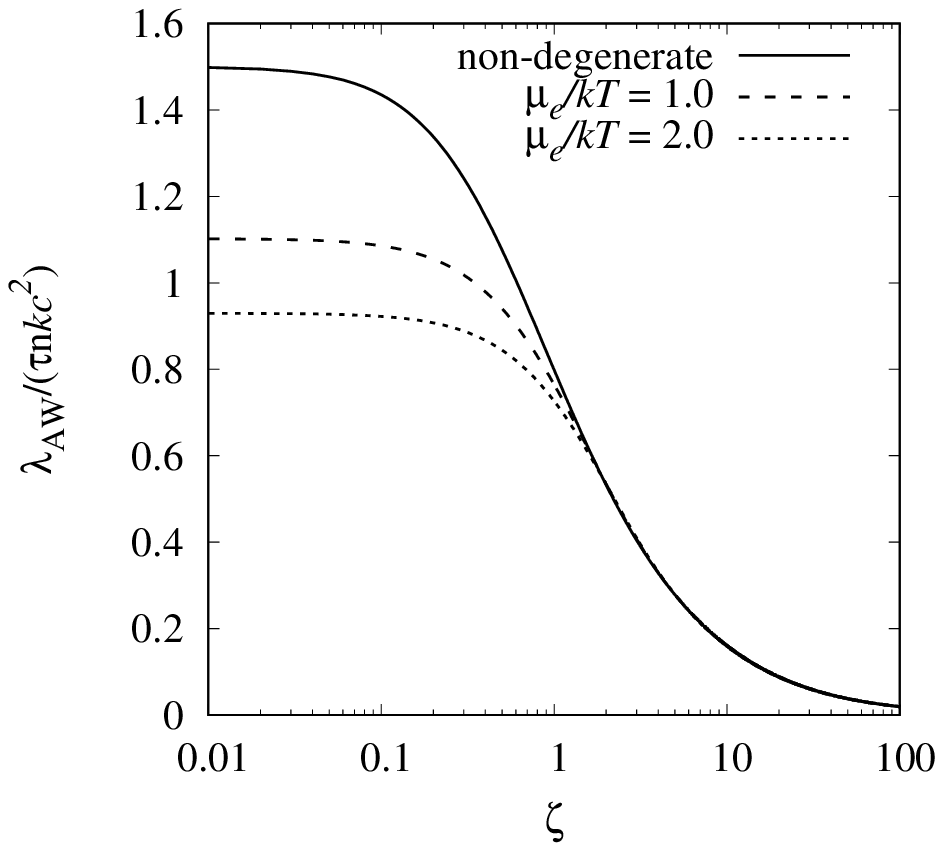}\includegraphics[width=6.1cm]{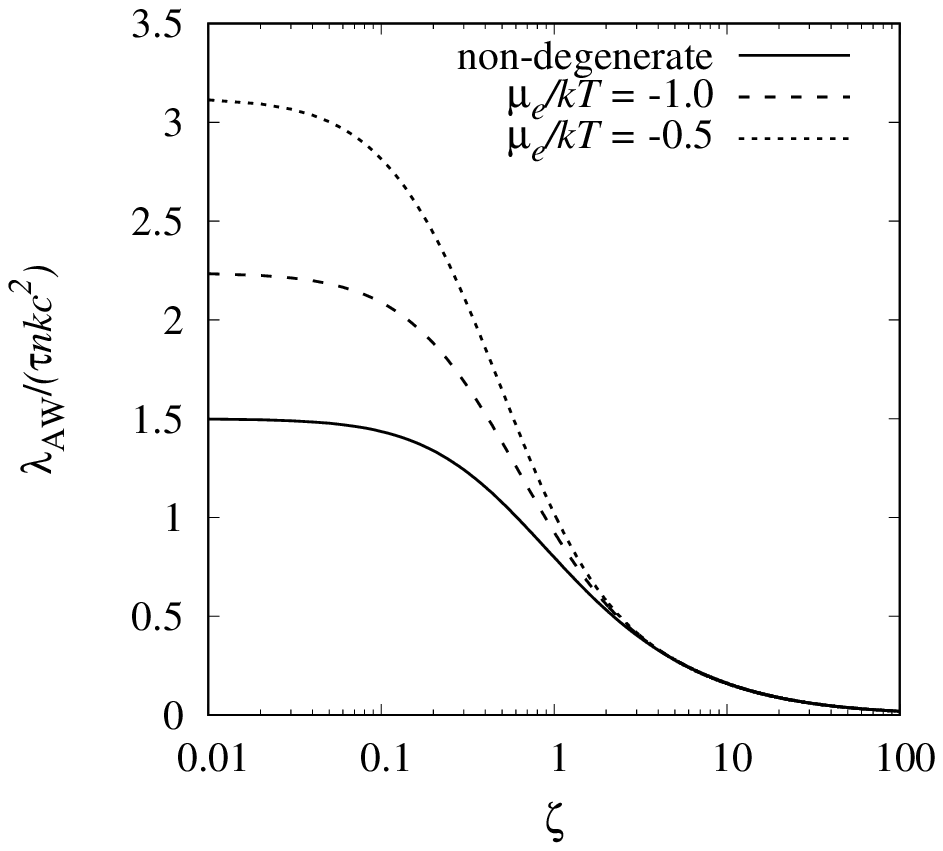}
\par\end{centering}
\begin{centering}
\includegraphics[width=6.1cm]{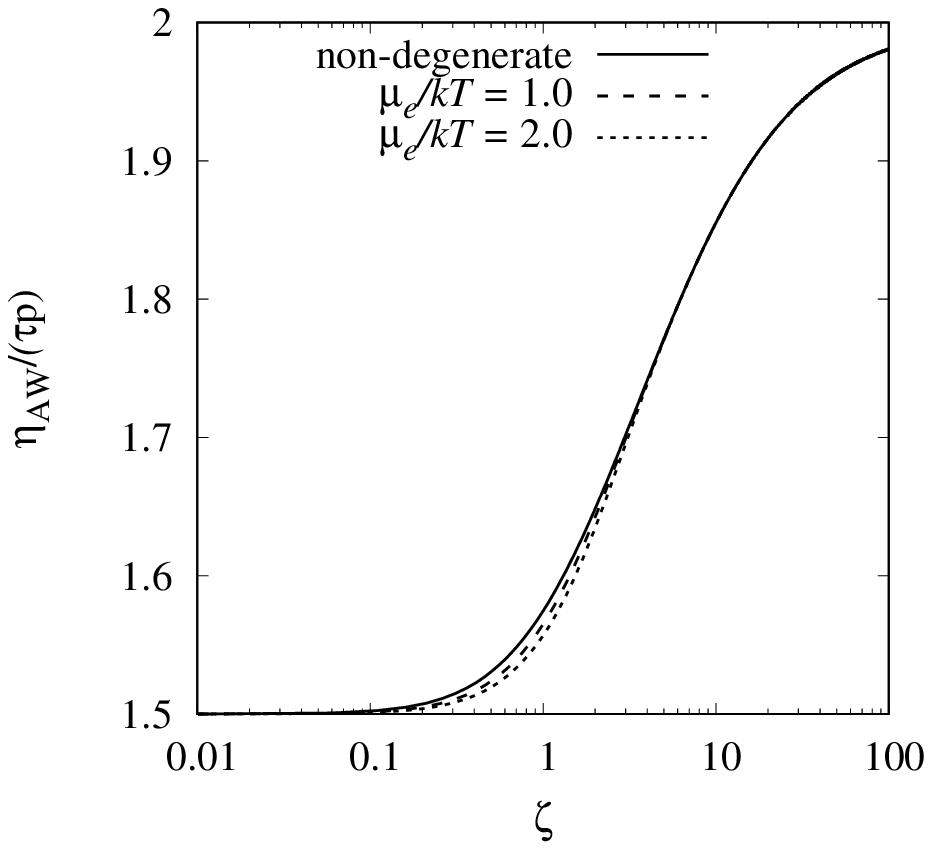}\includegraphics[width=6.1cm]{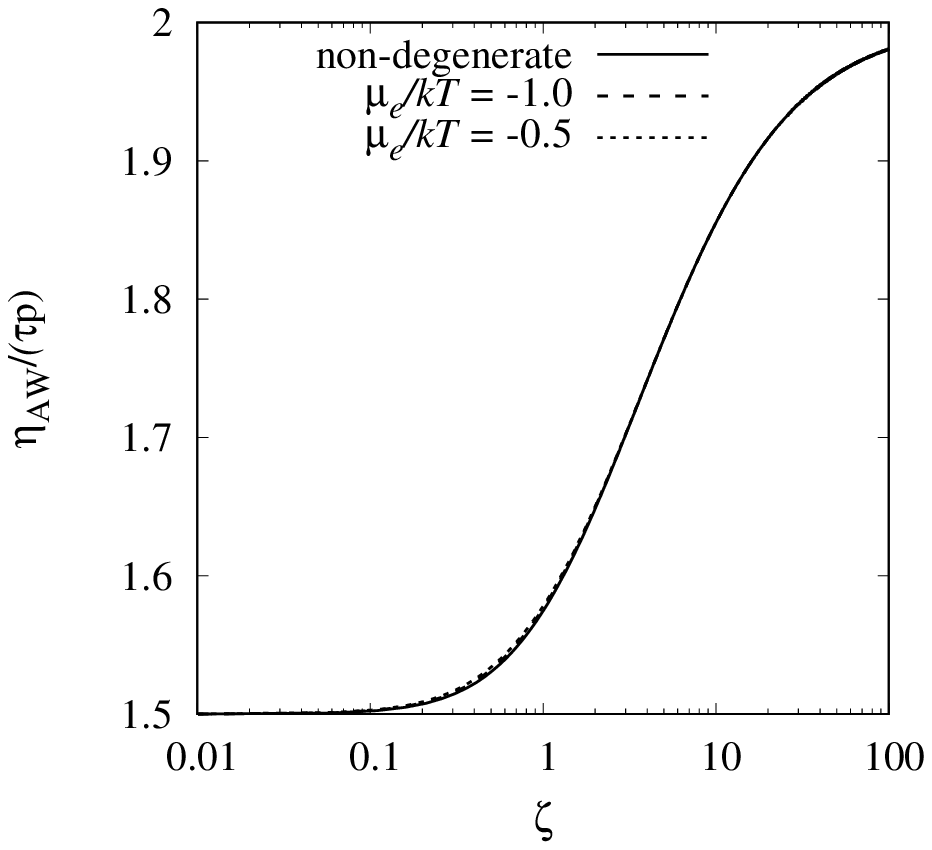}
\par\end{centering}
\begin{centering}
\includegraphics[width=6.1cm]{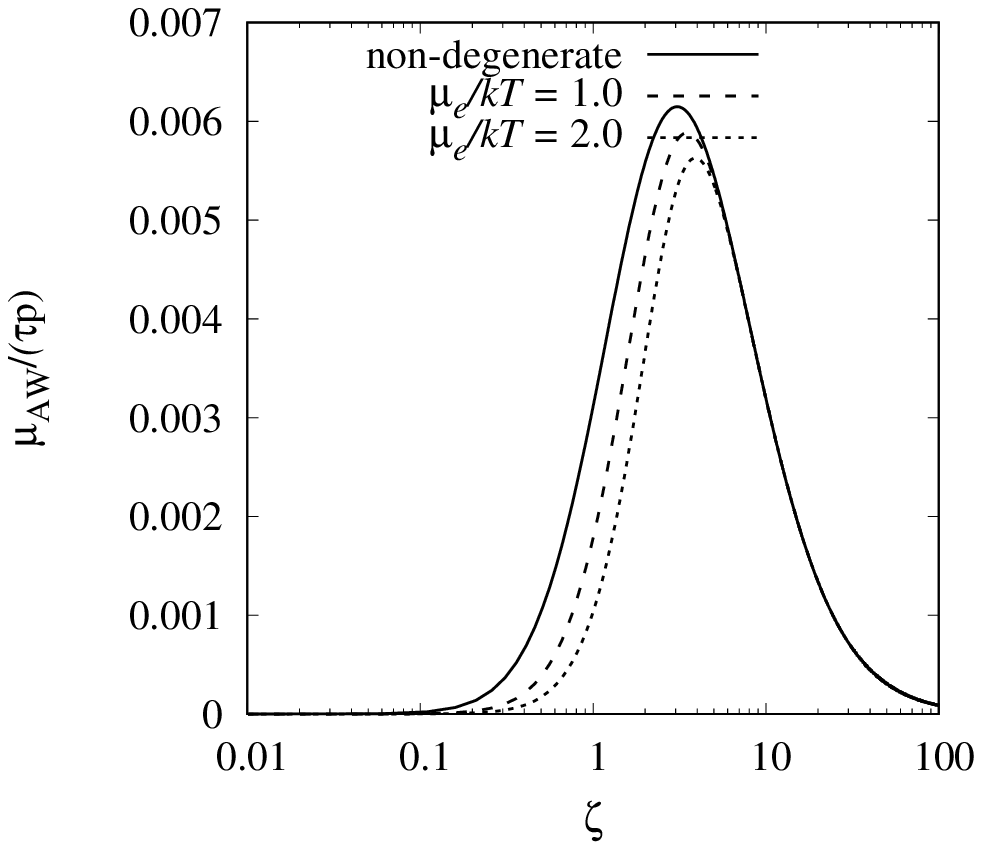}\includegraphics[width=6.1cm]{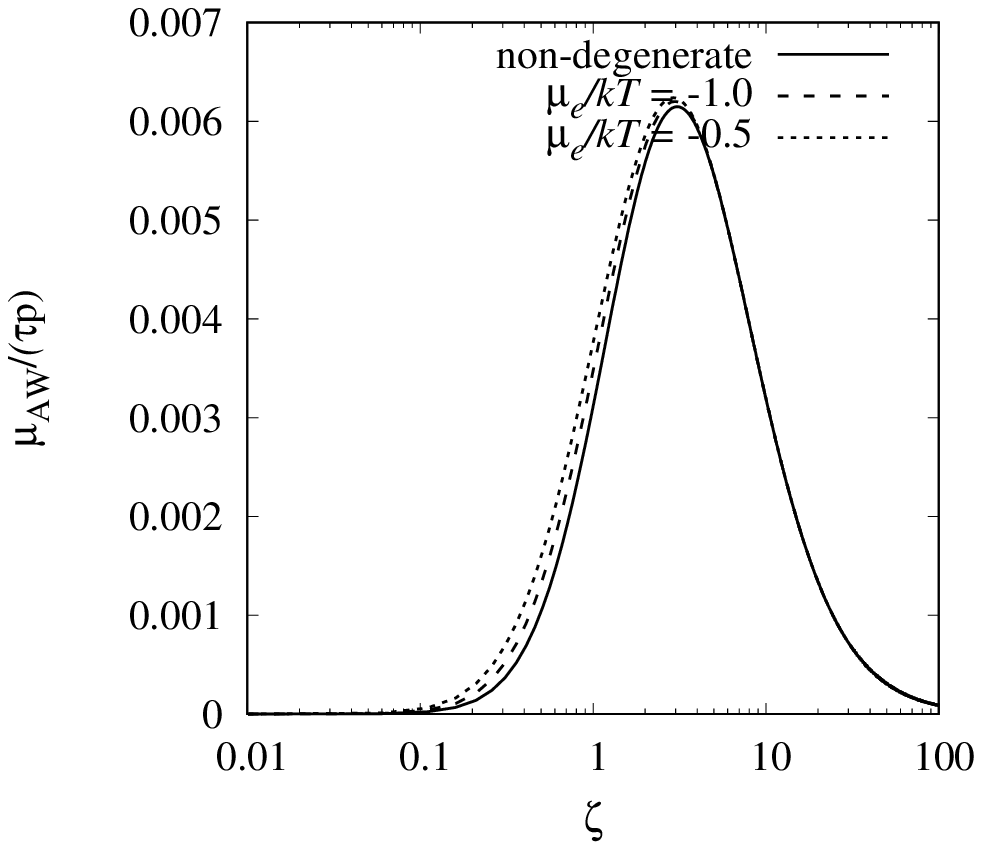}
\par\end{centering}
\caption{\label{fig:AW-1} The transport coefficients obtained with the Anderson-Witting
model: the thermal conductivity for fermions (top-left) and bosons
(top-right), the shear viscosity for fermions (center-left) and bosons
(center-right), and the bulk viscosity for fermions (bottom-left)
and bosons (bottom-right).}
\end{figure}

The establishment of the viscosity coefficients in this framework
follows directly from the substitution of $f_{{\rm \text{AWII}}}^{\left(1\right)}$
in Eq. (\ref{ji}). Following the same steps as those indicated in
Subsect. 3.4 one obtains the following expressions:
\begin{equation}
\eta_{{\rm AW}}=\tau_{{\rm AW}}\frac{\pi m^{3}c^{4}}{2}g_{s}\zeta\left(\mathcal{I}_{-1}^{s}-2\mathcal{I}_{1}^{s}+\mathcal{I}_{3}^{s}\right),\label{etaAW}
\end{equation}
\begin{align}
\mu_{{\rm AW}}&=-\tau_{{\rm AW}}g_{s}\pi m^{3}c^{4}\left\{ \left(\frac{\mathcal{I}_{2}^{s}I_{1}^{s}-\frac{1}{2}\mathcal{I}_{1}^{s}\left(3I_{2}^{s}-I_{0}^{s}
\right)}{\mathcal{I}_{1}^{s}\mathcal{I}_{3}^{s}-\left(\mathcal{I}_{2}^{s}\right)^{2}}
+\frac{\zeta}{2}\right)\left(\mathcal{I}_{1}^{s}-\mathcal{I}_{3}^{s}\right)\right. \nonumber\\
&-\left.\left(\frac{\mathcal{I}_{2}^{s}}{\mathcal{I}_{1}^{s}}\frac{\mathcal{I}_{2}^{s}I_{1}^{s}
-\frac{1}{2}\mathcal{I}_{1}^{s}\left(3I_{2}^{s}-I_{0}^{s}\right)}{\mathcal{I}_{1}^{s}
\mathcal{I}_{3}^{s}-\left(\mathcal{I}_{2}^{s}\right)^{2}}+\frac{I_{1}^{s}}{\mathcal{I}_{1}^{s}}
\right)\left(\mathcal{I}_{0}^{s}-\mathcal{I}_{2}^{s}\right)-\frac{\zeta}{2}
\left(\mathcal{I}_{-1}^{s}-\mathcal{I}_{1}^{s}\right) \right\} .\label{muAW}
\end{align}
These transport coefficients are plotted in Fig. \ref{fig:AW-1} for
$s=\pm1$.
For fermions, the three coefficients decrease as $ \mu_e $ increases. The bulk viscosity has a maximum as in the case of Marle's model, although the peaks are not at the same value. For bosons, thermal conductivity and shear viscosity increase with decreasing $|\mu_e / kT|$, while bulk viscosity decreases with increasing chemical potential in absolute value. In this case, both the thermal conductivity and bulk viscosity have the same behavior as the one shown in Fig. 1.  However, the shear viscosity tends to increase with $\zeta$, in agreement with previous results \cite{succi,Gabbana,Garcia-Mendez2019}, but opposite to Marle's corresponding coefficient.

\section{Non-degenerate limit}

In this section, we address the non-degenerate limit of the coefficients
obtained above, for which a comparison can be made with the results
of Appendix B for Marle's approximation and with the results of Gabbana,
et. al. in the Anderson-Witting calculation \cite{succi,Gabbana}. To assess this liming case in which the temperature is low but $\zeta^{-1}$ is still non-negligible, one can approximate both integrals $I_{n}^{s}$ and $\mathcal{I}_{n}^{s}$  by an exponential integral of negative order, for $n>0$, that is
\begin{equation}
\mathbb{I}_{n}\left(\zeta\right)=\frac{1}{h^{2}}\int_{1}^{\infty}
e^{-\zeta x}x^{n}dx=\frac{1}{h^{2}}\frac{e^{-\zeta}}{\zeta} \sum_{m=0}^{n}
\left(\frac{n!}{\left(n-m\right)!}\zeta^{-m}\right)\label{eq:fg}
\end{equation}
which satisfies
\begin{equation}
\zeta\mathbb{I}_{n}\left(\zeta\right)=n\mathbb{I}_{n-1}
+\frac{e^{-\zeta}}{h^{2}}\label{eq:rec}
\end{equation}
for $n\geq0$, and
\begin{equation}
\mathbb{I}_{-1}\left(\zeta\right)=\frac{1}{h^{2}}\int_{1}^{\infty}e^{-\zeta x}x^{-1}dx=\frac{1}{h^{2}}\Gamma\left(0,\zeta\right)\label{eq:gamma}
\end{equation}
for $n=-1$, where $\Gamma\left(0,\zeta\right)$ stands for the incomplete gamma function. In order to compare these limits, we introduce the expressions
above and the density $n$ in Eqs. (\ref{eq:lambda-Marle}), (\ref{5-1})
and (\ref{6-1}) such that Marle's transport coefficients in this
limit can be expressed as
\begin{equation}
\lambda_{\text{M}}=\tau_{\text M} nkc^{2}\frac{\zeta^{2}}{2}\left[\frac{1}{\zeta}
\left(1+\frac{\mathbb{I}_{3}}{\mathbb{I}_{1}}\right)-
\frac{\mathbb{I}_{0}}{\mathbb{I}_{1}}
\left(\frac{\mathbb{I}_{3}}{\mathbb{I}_{1}}-1\right)\right]
\end{equation}
\begin{equation}
\eta_{\text M}=\tau_{\text M}
nmc^{2}\left(\frac{\mathbb{I}_{3}}{\mathbb{I}_{1}}-1\right)\label{5-1-1}
\end{equation}
\begin{equation}
\mu_{\text M}=\frac{\tau_{\text M} nmc^{2}}{2} \left\{ \left(\frac{\mathbb{I}_{3}}{\mathbb{I}_{1}}-1\right)-
\frac{1}{\zeta}\frac{\left(\mathbb{I}_{1}\mathbb{I}_{4}-\mathbb{I}_{3}\mathbb{I}_{2}\right)}{\left(\mathbb{I}_{1}\mathbb{I}_{3}-\left(\mathbb{I}_{2}\right)^{2}\right)}
\right\} \label{6-1-1}
\end{equation}
which, using Eq. (\ref{eq:fg}) can be shown to reduce to Eqs. (\ref{eq:aq})-(\ref{aq2})
in Appendix B, that is their non-degenerate Marle-J\"uttner (MJ) value.

In the Anderson-Witting case, following the same steps as above one is lead to
\begin{equation}
\lambda_{{\rm AW}}=\tau_{\text AW} nkc^{2}\frac{\zeta^{2}}{2}\left(\frac{3}{\zeta}+\frac{\mathbb{I}_{0}}{\mathbb{I}_{1}}
\right)\left(-\frac{2}{\zeta}+\left(\frac{3}{\zeta}+\frac{\mathbb{I}_{0}}{\mathbb{I}_{1}}\right)\left(1-\frac{\mathbb{I}_{-1}}{\mathbb{I}_{1}}\right)\right),\label{as1}
\end{equation}
\begin{equation}
\eta_{\text AW}=\tau_{\text AW} \frac{nmc^{2}}{4}\zeta\left(\frac{\mathbb{I}_{-1}}{\mathbb{I}_{1}}+\frac{\mathbb{I}_{3}}{\mathbb{I}_{1}}-2\right),\label{as2}
\end{equation}
\begin{equation}
\mu_{\text AW}=-\tau_{\text AW}\frac{n m c^{2}}{2} \left\{ \frac{1}{2}\left(-\frac{2}{\zeta}\frac{\mathbb{I}_{1}}{\mathbb{I}_{1}\mathbb{I}_{3}-
\left(\mathbb{I}_{2}\right)^{2}}+\frac{\zeta}{\mathbb{I}_{1}}\right)
\left(\mathbb{I}_{1}-\mathbb{I}_{3}\right)+\frac{2}{\zeta}\left(-\frac{1}{\zeta}
\frac{\mathbb{I}_{1}\mathbb{I}_{2}}{\mathbb{I}_{1}\mathbb{I}_{3}-
\left(\mathbb{I}_{2}\right)^{2}}+1\right)-\frac{\zeta}{2}
\left(\frac{\mathbb{I}_{-1}}{\mathbb{I}_{1}}-1\right)\right\}, \label{as3}
\end{equation}
which, making use of Eqs. (\ref{eq:fg}) and (\ref{eq:gamma}), lead
to the following expressions

\begin{equation}
\lambda_{\text JAW}=\tau_{\text AW} nkc^{2}\frac{\zeta^{2}+3\zeta+3}{2\left(\zeta+1\right)^{3}}\left[\left(\zeta+1\right)
\left(\zeta^{2}+\zeta+1\right)-e^{\zeta}\zeta^{2}\left(\zeta^{2}+3\zeta+3\right)
\Gamma\left(0,\zeta\right)\right]\label{eq:aw1}
\end{equation}
\begin{equation}
\eta_{\text JAW}=\tau_{\text AW} \frac{nmc^{2}}{4\zeta}\left(\frac{-\zeta^{3}+\zeta^{2}+6\zeta+6+e^{\zeta}\zeta^{4}
\Gamma\left(0,\zeta\right)}{\zeta+1}\right)\label{aw2}
\end{equation}
\begin{equation}
\mu_{\text JAW}=\tau_{\text AW}
\frac{nmc^{2}}{4}\zeta^{3}\frac{e^{\zeta}\left(\zeta^{2}+4\zeta+2\right)\Gamma
\left(0,\zeta\right)-\zeta-3}{\left(\zeta+1\right)\left(\zeta^{2}+4\zeta+2\right)}
\label{aw3}
\end{equation}
which were obtained for the non-degenerate gas by Gabbana et. al. \cite{succi,Gabbana}, where the units $m=k=c=1$ were used.

Considering the low temperature, non-relativistic, scenario the transport
coefficients in Eqs. (\ref{as1})-(\ref{as3}) and (\ref{eq:aw1})-(\ref{aw3})
reduce to the 2D values obtained with the BGK approximation (see Appendix
C), that is
\begin{align*}
\lambda_{\text M} &\sim  \lambda_{\text MB} \left(1+\frac{1}{\zeta}-\dots \right),  &\lambda_{\text AW} &\sim  \lambda_{\text MB}\left(1-\frac{3}{\zeta}+\dots \right),\\
\eta_{\text M} &\sim  \eta_{\text MB}\left(1+\frac{2}{\zeta}+\frac{1}{\zeta^{2}}-\dots\right),  &\eta_{{\rm AW}} &\sim \eta_{\rm MB}\left(1-\frac{1}{\zeta}+\frac{4}{\zeta^{2}}+\dots \right),\\
\mu_{\text M} &\sim  -\frac{nkT\tau_{\text{M}}}{\zeta^{2}}\left(1-\frac{5}{\zeta}+\frac{19}{\zeta^{2}}-\dots
\right), &\mu_{{\rm AW}} &\sim \frac{nkT\tau_{\text{AW}}}{\zeta^{2}}\left(1-\frac{14}{\zeta}+
\frac{136}{\zeta^{2}}-\dots \right),
\end{align*}
where the subscript $\rm MB$ represents, as in Appendix C, Maxwell-Boltzmann's statistics. 
As in the three-dimensional case, the bulk viscosity vanishes in the limit $\zeta\rightarrow\infty$, which is consistent with the well known result of the non-relativistic ideal gas not presenting this type of dissipation (see Appendix C). 
It is worth noting that the dependence on $\zeta$ in this limit is similar for both models. For the comparison of the thermal conductivities, it is important
to note that in this limit, since $T/\left(nh_{e}\right)\sim1/\left(nk\zeta\right)$, we have
\[
\frac{\partial T}{\partial x^{d}}-\frac{T}{nh_{e}}\frac{\partial\text{p}}{\partial x^{d}}\sim\frac{\partial T}{\partial x^{d}}.
\]

It is also important to comment at this point that both sets of coefficients are proportional to a single relaxation time in this limit. As mentioned above, $\tau_{{\rm AW}}$ and $\tau_{{\rm M}}$ are relaxation parameters that can be modeled up to certain freedom, particularly the choice of average speed. Let us recall that this parameter must be of the same order of time between collisions, which in turn is proportional to the characteristic length scale divided by some average speed. The most suitable choice for $\tau$ depends on the system and the specific interactions studied. Possible options include M\o ller's velocity, mean velocity, average relative velocity, or the speed of sound \cite{KremerLibro,MendezAIP2010,Takamoto,nos19,Stauber,Cannoni}.
However, in the non-relativistic limit both parameters coincide and
are given by a single relaxation time which is precisely the one corresponding
to the BGK approximation $\tau$ as seen in Appendix C.

The opposite extreme of the temperature scale, the ultrarelativistic limit, can be assessed
by considering $\zeta\rightarrow0$. A somehow cumbersome procedure yields
\begin{align*}\lambda_{\text{M}} & \sim\frac{3c^{2}kn\tau_{\text{M}}}{\zeta}\left(1-\frac{\zeta}{3}+\dots\right), & \lambda_{\text{A}W} & \sim\frac{3c^{2}n k\tau_{\text{AW}}}{2}\left(1-\frac{4}{3}\zeta^{2}\left(1+\frac{9}{4}\gamma+\frac{9}{4}\ln\zeta\right)+\dots\right),\\
\eta_{\text{M}} & \sim\frac{6\text{p}\tau_{\text{M}}}{\zeta}\left(1+\frac{\zeta}{3}-\frac{\zeta^{2}}{3}+\dots\right), & \eta_{{\rm AW}} & \sim\frac{3\text{p}\tau_{\text{AW}}}{2}\left(1+\frac{1}{6}\zeta^{2}-\frac{1}{3}\zeta^{3}+\dots\right),\\
\mu_{\text{M}} & \sim-\frac{\text{p}\tau_{\text{M}}\zeta}{2}\left(1-3\zeta+\dots\right), & \mu_{{\rm AW}} & \sim\frac{3\text{p}\tau_{\text{AW}}\zeta^{2}}{8}\left(-\left[1+\frac{2}{3}\gamma+\frac{2}{3}\ln\zeta\right]+\frac{10}{3}\zeta-\dots\right),
\end{align*}
where $\gamma$ is Euler's constant. In this case one finds different behaviours
in the transport coefficients for each model. On the one hand, the ultrarelativistic limit of the transport coefficients obtained through Marle's approximation matches exactly the results obtained by Mendoza \textit{et al.}, in Ref. \cite{mendoza} for the same model. On the other hand, the results obtained using Anderson-Witting's frame deviate from the conductivity and viscosities calculated in Refs. \cite{succi,mendoza} only in a constant factor. However, in order to compare the coefficients calculated through relaxation models with those obtained with the full kernel, one has certain freedom in adjusting the relaxation parameters to match these limiting values. In this sense, Anderson-Witting's model is in general more likely to be appropriately matched with the results obtained by solving the complete kinetic equation while Marle's results present divergent quantities for which the matching is not straightforward. The analysis of the \rm{AW} transport coefficients in the ultrarelativistic non-degenerate case can be found in Ref. \cite{mendoza},  while the non-degenerate ultrarelativistic limit is handled in Ref. \cite{succi}.

\section{Final remarks}

In this work, a relaxation approximation has been applied to two-dimensional
degenerate systems together with the Chapman-Enskog approximation.
This type of approximation for the collisional relaxation to equilibrium
has proved successful in several scenarios and, as mentioned before,
has been shown to yield good numerical agreement among approximate
methods within kinetic theories in the systems here addressed. One
can show that in the relaxation approximation, the structure of the
constitutive equations is equivalent to the one obtained when the
complete kernel is considered. Two approximations have been considered, which differ in the direction of the temporal component for the 2+1 decomposition. Marle's model corresponds to the so-called particle frame while Landau's picture considers a frame in which heat dissipation is not included in the energy-momentum tensor and is thus referred to as the energy frame. The vectors chosen to determine each frame can be considered as almost parallel when the deviation from equilibrium in Landau's frame is negligible and the relative speed between both frames is non-relativistic.

Regarding the temperature dependence of the transport coefficients, good agreement is expected in the low and mild temperature scenarios, provided the relaxation parameters are modeled appropriately. The extreme non-degenerate ultrarelativistic case is well described in the Anderson-Witting framework while Marle's approximation requires an additional adjustment of such parameter, to agree with the results obtained by Anderson-Witting model and by the full relativistic Boltzmann equation \cite{MendezAIP2010,Takamoto,mendoza}. 

Even though Marle's model has been shown to appropriately describe thermal dissipation, the negative sign in the bulk viscosity clearly does not coincide with the Anderson-Witting case. The full Boltzmann results are only available for comparison in 2D for non-degenerate systems, in which case the bulk viscosity is also positive \cite{Garcia-Mendez2019}. It is worthwhile to point out that this sign disagreement seems not to be a peculiarity of the 2D case. At least, in the non-degenerate scenario, the same pathology is found for three dimensional systems. Indeed, as the reader can easily verify, the projection of the stress energy tensor yielding the bulk viscosity, considering the corresponding first order distribution function for this model (see Eq. (5.23) in Ref. \cite{KremerLibro}), also leads to a negative coefficient.
Alternative methods, including higher-order moments and a Maxwellian iteration procedure, seem to overcome this difficulty in the three-dimensional case \cite{KremerLibro}. For the two-dimensional case, more studies are necessary, but they go beyond the objectives of this work. At this point, it is fair to ask oneself why, being Marle's model unsatisfactory in some aspects, several works insist in its study. The answer, in the authors' point of view, is that the representation in which the space hypersurfaces are orthogonal to the hydrodynamic velocity make for a more physical description which also matches closely the non-relativistic counterpart of the theory.

The two sets of transport coefficients here obtained (Eqs. (\ref{eq:lambda-Marle}),
(\ref{5-1}), (\ref{6-1}) and (\ref{lambdaAW})-(\ref{muAW})) correspond
to the two approximations here invoked. Thus, it is important to keep
in mind that they are to be introduced within their corresponding
space-time decompositions in the hydrodynamic equations. It is worthwhile
at this point to emphasize that only when the ordering in dissipation of the equations
require it (measuring the deviation through the order in the gradients of state variables or through Knudsen's parameter), the particle and energy frames are to be considered equivalent. 
In general, the state variables do not coincide beyond first order
in an equilibrium parameter, a fact which needs to be carefully accounted
for. That is, the variables featured in the particle and energy-momentum fluxes in Eckart's and Landau's theory do not coincide in general and thus the transport coefficients introduced in them need to match the description invoked.

The results here obtained complete the dissipation picture for 2D
gases in the relaxation approximation for non-zero mass quantum particles
addressing the special relativistic degenerate case
as well as the non-degenerate low temperature gas, both in the particle and energy frames. The study of the transport properties in two dimensions for the zero
mass case within the kinetic theory framework is still an open problem
and will be addressed in the near future. Also, the threshold for
stability in the system of dissipative hydrodynamic equations for
the gases here studied is an interesting challenge . Indeed, by studying the linearized system of macroscopic equations, including the transport coefficients here calculated, one can obtain a dispersion relation which can be analyzed in order to determine the modified conditions for the corresponding fluctuations to be bounded. In such a case, predictions can be made on the properties of the Rayleigh-Brillouin spectrum which could, in principle, be measured. For details on such a calculation, the reader is referred to Ref. \cite{RB1}, where such modifications are addressed in the non-degenerate scenario.

\section*{Appendix A: The Euler Equations}

In order for $p^{a}\left(\partial f^{\left(0\right)}/\partial x^{a}\right)$
to be solely written in terms of the spatial gradients of the state
variables, one resorts to the Euler equations for the proper time
derivatives $u^{a}\left(\partial T/\partial x^{a}\right)$, $u^{a}\left(\partial\mu_{E}/\partial x^{a}\right)$
and $u^{b}\left(\partial u^{a}/\partial x^{b}\right)$. Such relations
are given by the local equilibrium transport equations for a dilute
system which are obtained from the conservation of the two tensor
quantities defined in Eqs. (\ref{eq:f}) and (\ref{f1}), in the local
equilibrium situation namely 
\begin{equation}
\frac{\partial N_{(0)}^{a}}{\partial x^{a}}=0\quad\qquad\text{and}\qquad\quad\frac{\partial T_{(0)}^{ab}}{\partial x^{b}}=0.\label{h}
\end{equation}
Notice that the particle flux as well as the energy-momentum tensor
have the same expressions both in the particle and energy frames.
In an equilibrium situation one has, considering $f=f^{\left(0\right)}$
in Eqs. (\ref{eq:f}) and (\ref{f1}): 
\begin{equation}
N_{(0)}^{a}=2\pi m^{2}c^{2}g_{s}u^{a}I_{1}^{s},
\end{equation}
\begin{equation}
T_{(0)}^{ab}=2\pi m^{3}c^{4}g_{s}\left(I_{2}^{s}\frac{u^{a}u^{b}}{c^{2}}+\frac{1}{2}\left(I_{0}^{s}-I_{2}^{s}\right)h^{ab}\right).
\end{equation}
Thus, the evolution equations in such regime are given by 
\begin{equation}
\frac{\partial}{\partial x^{a}}\left(u^{a}I_{1}^{s}\right)=0,\label{es}
\end{equation}
and 
\begin{equation}
\frac{\partial}{\partial x^{a}}\left(I_{2}^{s}\frac{u^{a}u^{b}}{c^{2}}+\frac{1}{2}\left(I_{0}^{s}-I_{2}^{s}\right)h^{ab}\right)=0,\label{e-1}
\end{equation}
where the space-time dependence is through the state variables $u^{\alpha}$,
$T$ and $\mu_{e}$. Notice that $u^{a}p_{a}=mc^{2}\left(1-p^{2}/\left(m^{2}c^{2}\right)\right)^{-1/2}$
is independent of $x^{a}$ ($p^{a}$ and $x^{a}$ as phase space variables
are independent). Using Eq. (\ref{eq:s-1}) one can show that the
following relation holds 
\begin{equation}
\frac{\partial I_{n}^{s}}{\partial x^{a}}=\frac{\zeta}{mc^{2}}\left(\left(mc^{2}\mathcal{I}_{n+1}^{s}-\mu_{e}\mathcal{I}_{n}^{s}\right)\frac{1}{T}\frac{\partial T}{\partial x^{a}}+\mathcal{I}_{n}^{s}\frac{\partial\mu_{e}}{\partial x^{a}}\right),\label{I}
\end{equation}
and thus, one can readily obtain from Eq. (\ref{es}) 
\begin{equation}
I_{1}^{s}\frac{\partial u^{a}}{\partial x^{a}}+\frac{\zeta u^{a}}{mc^{2}}\left[\frac{1}{T}\frac{\partial T}{\partial x^{a}}\left(mc^{2}\mathcal{I}_{2}^{s}-\mu_{e}\mathcal{I}_{1}^{s}\right)+\frac{\partial\mu_{e}}{\partial x^{a}}\mathcal{I}_{1}^{s}\right]=0.\label{a1}
\end{equation}
On the other hand, projecting Eq. (\ref{e-1}) in the direction parallel
and orthogonal to $u^{a}$ (the latter is obtained by contracting
with $h_{ab}$) and using Eq. (\ref{I}) one is lead to 
\begin{equation}
\left(3I_{2}^{s}-I_{0}^{s}\right)\frac{u^{b}}{c^{2}}\frac{\partial u^{a}}{\partial x^{b}}+\frac{\zeta}{mc^{2}}\left[\left(mc^{2}\left(\mathcal{I}_{1}^{s}-\mathcal{I}_{3}^{s}\right)+\mu_{e}\left(\mathcal{I}_{2}^{s}-\mathcal{I}_{0}^{s}\right)\right)\frac{1}{T}\frac{\partial T}{\partial x^{b}}+\left(\mathcal{I}_{0}^{s}-\mathcal{I}_{2}^{s}\right)\frac{\partial\mu_{e}}{\partial x^{b}}\right]h^{ab}=0\label{a2}
\end{equation}
and 
\begin{equation}
\left(mc^{2}\mathcal{I}_{3}^{s}-\mu_{e}\mathcal{I}_{2}^{s}\right)\frac{u^{a}}{T}\frac{\partial T}{\partial x^{a}}+\mathcal{I}_{2}^{s}u^{a}\frac{\partial\mu_{e}}{\partial x^{a}}=\frac{mc^{2}}{2\zeta}\left(I_{0}^{s}-3I_{2}^{s}\right)\frac{\partial u^{a}}{\partial x^{a}}\label{a3}
\end{equation}
respectively. Equations (\ref{a1})-(\ref{a3}) can be solved for
the proper time derivatives $u^{a}\frac{\partial T}{\partial x^{a}}$,
$u^{a}\frac{\partial\mu_{e}}{\partial x^{a}}$ and $u^{a}\frac{\partial u^{b}}{\partial x^{a}}$.
The calculation is straightforward and yields: 
\begin{equation}
\left(\frac{u^{a}}{T}\frac{\partial T}{\partial x^{a}}\right)=\frac{\mathcal{I}_{2}^{s}I_{1}^{s}+\frac{1}{2}\mathcal{I}_{1}^{s}\left(I_{0}^{s}-3I_{2}^{s}\right)}{\mathcal{I}_{1}^{s}\mathcal{I}_{3}^{s}-\left(\mathcal{I}_{2}^{s}\right)^{2}}\frac{1}{\zeta}\frac{\partial u^{a}}{\partial x^{a}}\label{tpunto}
\end{equation}
\begin{equation}
\left(u^{a}\frac{\partial\mu_{e}}{\partial x^{a}}\right)=-\left\{ I_{1}^{s}-\left(\mathcal{I}_{2}^{s}-\alpha\mathcal{I}_{1}^{s}\right)\left(\frac{\mathcal{I}_{2}^{s}I_{1}^{s}+\frac{1}{2}\mathcal{I}_{1}^{s}\left(I_{0}^{s}-3I_{2}^{s}\right)}{\mathcal{I}_{1}^{s}\mathcal{I}_{3}^{s}-\left(\mathcal{I}_{2}^{s}\right)^{2}}\right)\right\} \frac{mc^{2}}{\zeta}\frac{1}{\mathcal{I}_{1}}\frac{\partial u^{a}}{\partial x^{a}}\label{mupunto}
\end{equation}
\begin{equation}
\left(u^{b}\frac{\partial u^{a}}{\partial x^{b}}\right)=-\frac{\zeta{c^{2}}}{\left(3I_{2}^{s}-I_{0}^{s}\right)}h^{ab}\left\{ \left(\mathcal{I}_{0}^{s}-\mathcal{I}_{2}^{s}\right)\frac{\alpha}{\mu_{e}}\frac{\partial\mu_{e}}{\partial x^{b}}+\left(\left(\mathcal{I}_{1}^{s}-\mathcal{I}_{3}^{s}\right)+\alpha\left(\mathcal{I}_{2}^{s}-\mathcal{I}_{0}^{s}\right)\right)\frac{1}{T}\frac{\partial T}{\partial x^{b}}\right\} \label{upunto}
\end{equation}
Equations (\ref{tpunto}) and (\ref{mupunto}) are Euler equations
used to obtain Eq. (\ref{pb}). 

\section*{Appendix B: The non-degenerate special relativistic case}

This section addresses the non-degenerate limit of the calculation carried out in the main article and follows basically the same steps. However, only Marle's approximation is here considered since the calculation in the Anderson-Witting case was already addressed by Gabbana et al. and can be found in \textcolor{teal}{Refs. \cite{succi,Gabbana}}. In order to perform
such task, one starts by considering a two-dimensional Maxwell-J\"uttner
distribution, that is 
\begin{equation}
f^{\left(0\right)}=\frac{ne^{\zeta}\zeta^{2}}{2\pi m^{2}c^{2}\left(\zeta+1\right)}\exp\left(-\zeta x\right)\label{a}
\end{equation}
from which one can readily obtain for the internal energy
\[
\epsilon=mc^{2}\left(\frac{2+\zeta\left(2+\zeta\right)}{\zeta\left(\zeta+1\right)}\right)
\]
and ${\rm p}=nkT$ for the hydrostatic pressure. For the right hand
sides of Eqs. (\ref{marle}) and (\ref{aw}) one obtains
\begin{align}
p^{a}\frac{\partial f^{\left(0\right)}}{\partial x^{a}} & =f^{\left(0\right)}\left\{ h^{ab}p_{b}\left[-\left(\frac{\epsilon}{mc^{2}}-x\right)\frac{\zeta}{T}\frac{\partial T}{\partial x^{a}}+\frac{1}{n}\frac{\partial n}{\partial x^{a}}-\zeta\frac{p_{b}}{mc^{2}}\frac{\partial u^{b}}{\partial x^{a}}\right]\right.\nonumber \\
 & +\left.mx\left[-\left(\frac{\epsilon}{mc^{2}}-x\right)\frac{\zeta}{T}\left(u^{a}\frac{\partial T}{\partial x^{a}}\right)+\frac{1}{n}\left(u^{a}\frac{\partial n}{\partial x^{a}}\right)-\zeta\frac{p_{b}}{mc^{2}}\left(u^{a}\frac{\partial u^{b}}{\partial x^{a}}\right)\right]\right\} \label{cx}
\end{align}
where time and space derivatives have been separated by considering
$p^{a}=h^{ab}p_{b}+mxu^{a}$ as before. The local equilibrium equations
are obtained form Eq. (\ref{h}) with $N_{\left(0\right)}^{a}$ and
$T_{\left(0\right)}^{ab}$ given by
\begin{equation}
N_{\left(0\right)}^{a}=nu^{a},\qquad T_{\left(0\right)}^{ab}=\frac{n\epsilon}{c^{2}}u^{a}u^{b}-\mathrm{p}h^{ab},\label{n-1-2}
\end{equation}
which leads to
\begin{equation}
u^{a}\frac{\partial n^{a}}{\partial x^{a}}=-n\frac{\partial u^{a}}{\partial x^{a}},\label{eq:s-2}
\end{equation}
\begin{equation}
u^{a}\frac{\partial T}{\partial x^{a}}=-\frac{mc^{2}}{c_{n}\zeta}\frac{\partial u^{a}}{\partial x^{a}},\label{eq:t-1}
\end{equation}
\begin{equation}
u^{a}\frac{\partial u^{d}}{\partial x^{a}}=\frac{c^{2}}{nh_{e}}h^{ad}\left(\frac{p}{n}\frac{\partial n}{\partial x^{a}}+\frac{p}{T}\frac{\partial T}{\partial x^{a}}\right),\label{eq:u-1}
\end{equation}
 where $h_{e}=\epsilon+kT$ and
\[
c_{n}=k\frac{\zeta^{2}+4\zeta+2}{\left(\zeta+1\right)^{2}}.
\]
Substitution of Eqs. (\ref{eq:s-2})-(\ref{eq:u-1}) in Eq. (\ref{cx})
leads to
\begin{align}
p^{a}\frac{\partial f^{\left(0\right)}}{\partial x^{a}} & =f^{\left(0\right)}\left\{ h^{ab}p_{b}\left(\frac{mc^{2}}{h_{e}}x-1\right)\left[\frac{\epsilon\zeta}{mc^{2}}\left(\frac{1}{T}\frac{\partial T}{\partial x^{a}}\right)-\left(\frac{1}{n}\frac{\partial n}{\partial x^{a}}\right)\right]\right.\nonumber \\
 & -\left.mx\left(1+\frac{k}{c_{n}}\zeta\left(x-\frac{\epsilon}{mc^{2}}\right)\right)\left(\frac{\partial u^{a}}{\partial x^{a}}\right)-\frac{\zeta}{mc^{2}}h^{ab}p_{b}p^{c}\frac{\partial u_{c}}{\partial x^{a}}\right\} ,\label{e}
\end{align}
where the first line corresponds to the thermal dissipation contribution
while the second line contains the viscous terms, in which we once
again introduce the usual decomposition for the velocity
gradient mentioned in Sect. 2.2. Considering Eq. (\ref{marle}) one
obtains $f_{{\rm JM}}^{\left(1\right)}=f_{{\rm JMI}}^{\left(1\right)}+f_{{\rm JMII}}^{\left(1\right)}$
(where ${\rm JM}$ refers to J\"uttner-Marle), with
\[
f_{{\rm JMI}}^{\left(1\right)}=-\frac{\tau}{m}f^{\left(0\right)}h^{ed}p_{d}\left(\frac{mc^{2}}{h_{e}}x-1\right)\left[\frac{\epsilon\zeta}{mc^{2}}\left(\frac{1}{T}\frac{\partial T}{\partial x^{a}}\right)-\left(\frac{1}{n}\frac{\partial n}{\partial x^{a}}\right)\right],
\]
\[
f_{{\rm JMII}}^{\left(1\right)}=\tau f^{\left(0\right)}\left[\frac{\zeta}{m^{2}c^{2}}h_{b}^{a}p^{b}p^{c}\mathring{\sigma}_{ac}-\left(\left(\frac{1}{2}-\frac{k}{c_{n}}\right)\zeta x^{2}+\left(\frac{k}{c_{n}}\frac{\epsilon\zeta}{mc^{2}}-1\right)x-\frac{\zeta}{2}\right)\left(\frac{\partial u^{a}}{\partial x^{a}}\right)\right].
\]

Considering $f_{{\rm JMI}}^{\left(1\right)}$ in Eq. (\ref{j}) and
using once again Eq. (\ref{d-1}) one is lead to

\[
q^{a}=-\tau h^{ae}\left[\frac{\epsilon\zeta}{mc^{2}}\left(\frac{1}{T}\frac{\partial T}{\partial x^{a}}\right)-\left(\frac{1}{n}\frac{\partial n}{\partial x^{a}}\right)\right]\frac{ne^{\zeta}\zeta^{2}}{\left(\zeta+1\right)}\frac{mc^{4}}{2}\int\exp\left(-\zeta x\right)\left(\frac{mc^{2}}{h_{e}}x^{2}-x\right)\left(1-x^{2}\right)dx,
\]
which leads to
\[
q^{a}=h^{ad}\frac{nmc^{4}}{2}\frac{\left(6+4\zeta\left(3+\zeta\right)\right)}{\zeta^{2}\left(\zeta+1\right)\left(\zeta^{2}+3\zeta+3\right)}\left[\left(\frac{h_{e}}{kT}-1\right)\left(\frac{1}{T}\frac{\partial T}{\partial x^{d}}\right)-\left(\frac{1}{n}\frac{\partial n}{\partial x^{d}}\right)\right]
\]

or
\[
q^{a}=\tau nkc^{2}\frac{\left(3+2\zeta\left(3+\zeta\right)\right)}{\zeta\left(\zeta+1\right)^{2}}h^{ad}\left[\left(\frac{\partial T}{\partial x^{d}}\right)-\frac{T}{nh_{e}}\left(\frac{\partial p}{\partial x^{d}}\right)\right]
\]
and thus
\begin{equation}
\lambda_{{\rm JM}}=\tau nc^{2}k\frac{\left(3+2\zeta\left(3+\zeta\right)\right)}{\zeta\left(\zeta+1\right)^{2}}\label{eq:aq}
\end{equation}
For the viscous component of the energy-momentum tensor one has
\begin{align*}
\Pi^{mn} & =h_{i}^{m}h_{j}^{n}\int p^{i}p^{j}f_{{\rm JMII}}^{\left(1\right)}\left(\frac{d^{2}p}{p_{0}}c\right)
\end{align*}
which, using the properties mentioned in the main text, leads to the
following expression 
\begin{equation}
h_{c}^{a}h_{d}^{b}\int A\left(x\right)p^{c}p^{d}dx=\frac{m^{2}c^{2}}{2}h^{ab}\int A\left(x\right)\left(1-x^{2}\right)dx,\nonumber 
\end{equation}
\begin{align*}
\Pi^{mn} & =\tau n\frac{mc^{2}e^{\zeta}\zeta^{3}}{4\left(\zeta+1\right)}\left\{ \mathring{\sigma}^{mn}\int\exp\left(-\zeta x\right)\left(x^{2}-1\right)^{2}dx\right.\\
 & \left.+h^{mn}\left(\frac{\partial u^{a}}{\partial x^{a}}\right)\int\exp\left(-\zeta x\right)\left(\frac{\zeta^{2}x^{2}-2\zeta\left(\zeta+2\right)x}{\zeta^{2}+4\zeta+2}+1\right)\left(1-x^{2}\right)dx\right\} 
\end{align*}
from which one can readily obtain the viscous coefficients for the
high temperature non-degenerate gas within Marle's model: 
\begin{equation}
\eta_{{\rm JM}}=2\tau nkT\frac{\left(\zeta^{2}+3\zeta+3\right)}{\zeta\left(\zeta+1\right)}\label{aq1}
\end{equation}
\begin{equation}
\mu_{{\rm JM}}=-\tau nkT\frac{\zeta}{\left(\zeta+1\right)\left(\zeta^{2}+4\zeta+2\right)}\label{aq2}
\end{equation}
Notice that, in the non-relativistic limit

\[
\lambda_{{\rm JM}}\sim\frac{2c^{2}nk\tau}{\zeta}\left(1+\frac{1}{\zeta}+\dots\right)
\]

\[
\eta_{{\rm JM}}\sim2nkT\tau\left(1+\frac{2}{\zeta}+\dots\right)
\]
\[
\mu_{{\rm JM}}\sim-nkT\tau\frac{1}{\zeta^{2}}\left(1-\frac{5}{\zeta}+\dots\right).
\]

\section*{Appendix C: The 2D BGK approximation for the non-degenerate and non-relativistic
scenario.}

In this appendix, the determination for the relaxation approximation
for the classical non relativistic transport coefficients is briefly
outlined. Also, the fact that the low temperature ideal gas has vanishing
bulk viscosity is pointed out. The starting point for such calculation
is the relaxation time BGK approximation for the Boltzmann equation
namely \cite{BGK}, 
\[
\frac{\partial f}{\partial t}+\vec{v}\cdot\frac{\partial f}{\partial\vec{r}}=-\frac{f-f^{\left(0\right)}}{\tau}
\]
where we have assumed no external potentials are present. Here $\vec{v}$
stands for the molecular vector velocity and $\tau$ has the same
interpretation as in the main text, being in the non-relativistic
case clearly associated with the characteristic time for the distribution
function to relax to its local equilibrium form, which in this 2D case
is given by (Ref. \cite{HFBidi}) 
\[
f^{\left(0\right)}=\frac{nm}{2\pi kT}e^{-\frac{mC^{2}}{2kT}}
\]
The so-called chaotic or peculiar velocity, namely the molecule's
velocity measured by an observer co-moving with the fluid, is denoted
by $\vec{C}=\vec{v}-\vec{u}$ with $C$ being its magnitude and $\vec{u}$
the system's hydrodynamic velocity:

\[
\vec{u}=\frac{1}{n}\int f^{\left(0\right)}\vec{v}d^{2}v
\]
Introducing the Chapman-Enskog first order expansion (see Eq. (\ref{eq:ce}))
one can obtain the following expression for the corresponding first
order correction to $f^{\left(0\right)}$: 
\[
f^{\left(1\right)}=-\tau f^{(0)}\left\{ \vec{C}\cdot\left[\left(\frac{mC^{2}}{2kT}-1\right)\frac{\nabla T}{T}+\frac{\nabla n}{n}+\frac{m\vec{C}}{kT}.\nabla\vec{u}\right]+\frac{m\vec{C}}{kT}\cdot\frac{d\vec{u}}{dt}+\left(\frac{mC^{2}}{2kT}-1\right)\frac{1}{T}\frac{dT}{dt}+\frac{1}{n}\frac{dn}{dt}\right\} 
\]
The introduction of the Euler equations in this limit namely,
\[
\frac{dn}{dt}=-n\nabla\cdot\vec{u},\qquad\quad\frac{dT}{dt}=-T\nabla\cdot\vec{u},\qquad\quad\frac{d\vec{u}}{dt}=-\frac{\nabla{\rm p}}{mn}
\]
where $d/dt=\partial/\partial t+\vec{u}\cdot\nabla$
leads to
\[
f^{\left(1\right)}=-\tau f^{(0)}\left\{ \vec{C}\cdot\left[\left(\frac{mC^{2}}{2kT}-2\right)\frac{\nabla T}{T}+\frac{m\vec{C}}{kT}\nabla\vec{u}\right]-\left(\frac{mC^{2}}{2kT}\right)\nabla\cdot\vec{u}\right\} 
\]
The dissipative fluxes for the Maxwell-Boltzmann case are given by
the following averages,
\[
\vec{Q}=\int\left(\frac{m}{2}C^{2}\right)\vec{C}f^{\left(1\right)}d^{2}v
\]
\[
\overleftrightarrow{\Pi}=m\int\vec{C}\vec{C}f^{\left(1\right)}d^{2}v
\]
One obtains for the heat flux:
\begin{equation}
\vec{Q}=-\tau\int\left(\frac{m}{2}C^{2}\right)\vec{C}\left[\left(\frac{mC^{2}}{2kT}-2\right)\vec{C}\cdot\frac{\nabla T}{T}\right]f^{(0)}d^{2}v\label{eq:f-1}
\end{equation}
 which, using
\[
\int A\left(C\right)\vec{C}\vec{C}d^{2}v=2\pi\int A\left(C\right)\vec{C}\vec{C}CdC=\mathbb{I}\pi\int A\left(C\right)C^{3}dC
\]
leads to
\[
\vec{Q}=-\tau\frac{2n\left(kT\right)^{2}}{m}\frac{\nabla T}{T}
\]
Thus in this (Maxwell-Boltzmann) limit $\lambda_{{\rm MB}}=\tau\frac{2n\left(kT\right)^{2}}{m}$.

For the stress tensor one has

\begin{equation}
\overleftrightarrow{\Pi}=-\tau m\int\vec{C}\vec{C}f^{(0)}\frac{m}{kT}\left(\vec{C}\vec{C}:\nabla\vec{u}-\frac{C^{2}}{2}\nabla\cdot\vec{u}\right)d^{2}v\label{be}
\end{equation}
It is worthwhile to put special attention to the following step. Even
though it is widely known that the non-relativistic ideal gas does not oppose to compression or expansion stresses, which is usually referred to as having a vanishing bulk viscosity, rarely it is clearly understood why this is so. From the microscopic point of view, one can clearly
identify that when averaging momentum exchange, the compressibility
does not play a role. Indeed, since the velocity gradient can be written
as 
\begin{equation}
\nabla\vec{u}=\mathring{\overleftrightarrow{\sigma}}+\overleftrightarrow{w}+\frac{1}{2}\mathbb{I}\nabla\cdot\vec{u}\label{eq:a-2}
\end{equation}
where $\mathring{\overleftrightarrow{\sigma}}$ is the symmetric traceless
part of the velocity gradient and $\overleftrightarrow{w}$ its antisymmetric
component. The latter vanishes when contracted with the symmetric
dyad $\vec{C}\vec{C}$. Introducing Eq. (\ref{eq:a-2}) in Eq. (\ref{pb})
and writing $\vec{C}\vec{C}=\mathring{\overleftrightarrow{CC}}+\frac{1}{2}C^{2}\mathbb{I}$,
one is lead to the mentioned result: momentum dissipation in ideal
non-relativistic gases is solely driven by shear stresses which can
be also understood as a vanishing bulk viscosity. However it is important
to bear in mind that it is not that such a coefficient is zero in
the microscopic formalism but that instead, the compression stresses
do not appear at all in the stress tensor.

Carrying out the steps outlined in the paragraph above, the constitutive
equation for the Navier tensor can be written as 
\[
\overleftrightarrow{\Pi}=-\tau\frac{nm^{3}}{2\pi\left(kT\right)^{2}}\mathring{\overleftrightarrow{\sigma}}:\int\vec{C}\vec{C}\vec{C}\vec{C}e^{-\frac{mC^{2}}{2kT}}d^{2}v
\]
\begin{equation}
\overleftrightarrow{\Pi}=-\tau\frac{m^{2}}{kT}\frac{nm}{2\pi kT}\int\vec{C}\vec{C}\left(\mathring{\overleftrightarrow{CC}}:\mathring{\overleftrightarrow{\sigma}}\right)e^{-\frac{mC^{2}}{2kT}}d^{2}v\label{intNR}
\end{equation}
The integral in Eq. (\ref{intNR}) can be performed for each component
and leads to the following constitutive equation: 
\begin{equation}
\overleftrightarrow{\Pi}=-2nkT\tau\mathring{\overleftrightarrow{\sigma}}.\label{StrTensorNR}
\end{equation}

Thus, for the Maxwell-Boltzmann case one obtains
\begin{equation}
\eta_{{\rm MB}}=2nkT\tau
\end{equation}
and
\begin{equation}
\mu_{{\rm MB}}=0.
\end{equation}

\end{document}